%% file: manuscript.tex
\documentclass[aps,prl,superscriptaddress,amsmath,amssymb,floatfix,reprint]{revtex4-2}
\usepackage{graphicx}
\usepackage{dcolumn}
\usepackage{bm}
\usepackage{physics}
\usepackage{braket} 
\usepackage{color}
\usepackage[colorlinks=true, citecolor=blue, linkcolor=blue, urlcolor=blue]{hyperref}
\usepackage[T1]{fontenc}
\usepackage{tgtermes}

\graphicspath{ {./}{./figs/} }

\begin{document}
\title{Net and Hidden Spin-Valley Locking Enable Ultrahigh Hole Mobility in Covalent Bulk WN$_2$}

\author{Rong-Tian Pang}
\thanks{These authors contributed equally to this work}
\affiliation{Centre for Quantum Physics, Key Laboratory of Advanced Optoelectronic Quantum Architecture and Measurement (MOE), School of Physics, Beijing Institute of Technology, Beijing 100081, China.}

\author{Zhongjuan Han}
\thanks{These authors contributed equally to this work}
\affiliation{Key Laboratory of Advanced Materials and Devices for Post-Moore Chips, Ministry of Education, University of Science and Technology Beijing, Beijing 100083, China}
\affiliation{School of Mathematics and Physics, University of Science and Technology Beijing, Beijing 100083, China}

\author{Jiayi Gong}
\affiliation{Centre for Quantum Physics, Key Laboratory of Advanced Optoelectronic Quantum Architecture and Measurement (MOE), School of Physics, Beijing Institute of Technology, Beijing 100081, China.}

\author{Jiangang He}
\email{jghe2021@ustb.edu.cn}
\affiliation{Key Laboratory of Advanced Materials and Devices for Post-Moore Chips, Ministry of Education, University of Science and Technology Beijing, Beijing 100083, China}
\affiliation{School of Mathematics and Physics, University of Science and Technology Beijing, Beijing 100083, China}

\author{Jin-Jian Zhou}
\email{jjzhou@bit.edu.cn}
\affiliation{Centre for Quantum Physics, Key Laboratory of Advanced Optoelectronic Quantum Architecture and Measurement (MOE), School of Physics, Beijing Institute of Technology, Beijing 100081, China.}
\affiliation{International Center for Quantum Materials, Beijing Institute of Technology, Zhuhai, 519000, China.}

\author{Yugui Yao}
\affiliation{Centre for Quantum Physics, Key Laboratory of Advanced Optoelectronic Quantum Architecture and Measurement (MOE), School of Physics, Beijing Institute of Technology, Beijing 100081, China.}
\affiliation{International Center for Quantum Materials, Beijing Institute of Technology, Zhuhai, 519000, China.}

\date{\today}

\begin{abstract}
High carrier mobility at room temperature underpins high-performance electronics, yet high hole mobility remains rare in bulk semiconductors.
Spin-valley locking can suppress intervalley scattering and enhance mobility, but it is limited to materials with broken inversion symmetry.
Hidden spin polarization offers a possible route beyond this constraint, although whether its compensated spin textures could protect charge transport remains unclear.
Using \textit{ab initio} electron-phonon and transport calculations, we show that the two hexagonal phases of bulk WN$_2$ realize net and hidden spin-valley locking and exhibit 
ultrahigh room-temperature hole mobilities.
In non-centrosymmetric $\alpha$-WN$_2$, a large valley spin splitting produces net spin-valley locking that nearly eliminates phonon-mediated intervalley scattering.
In centrosymmetric $\beta$-WN$_2$, hidden Zeeman-type spin polarization yields a compensated, sector-resolved spin texture that reverses between valleys and suppresses intervalley scattering as effectively as the net locking does.
The stiff W--N/N--N covalent network further keeps the remaining intravalley scattering weak.
Our results establish hidden spin polarization as an effective transport-protection mechanism and extend spin-valley engineering to centrosymmetric bulk semiconductors.
\end{abstract}
\maketitle

\label{introduction}
\input{intro.tex}

\label{result-1}
\input{result-1.tex}

\label{result-2}
\input{result-2.tex}

\label{result-3}
\input{result-3.tex}


\label{conclusion}
\input{conclusion.tex}

\begin{acknowledgments}
\textit{Acknowledgments}---R.-T. P and J.-J. Z acknowledge financial support from the National Natural Science Foundation of China (Grant Nos.~12574250, 12104039), the Beijing Natural Science Foundation (Grant No.~Z260002),  and the National Key R\&D Program of China (Grant No.~2022YFA1403400). Z.H. and J.H. acknowledge the support of the National Natural Science Foundation of China (Grant No.~12374024) and the Fundamental Research Funds for the Central Universities (No.~FRF-BRA-26-007)
\end{acknowledgments}

\section*{Data Availability}
The data that support the findings of this article are openly available~\cite{data}.

\bibliography{wn2-refs}
\end{document}

%% file: intro.tex

The pursuit of high-performance, energy-efficient electronics relies on semiconductors with both high carrier mobility and sizable band gap, a combination that remains elusive, particularly for holes~\cite{DelAlamo2011NanometreScaleElectronics,Hautier2013IdentificationDesignPrinciples}.
Valence-band edges are typically heavy and nearly degenerate, opening a large phase space for phonon scattering that limits hole mobility~\cite{Gibbs2017EffectiveMass}.
GaAs has a high electron mobility approaching 9000~cm$^2$/Vs at room temperature, yet its hole mobility is much lower, about 400~cm$^2$/Vs~\cite{Blakemore1982Semiconducting,Wolfe1970ElectronMobility,Zhou2016AbInitio,Liu2017GaAs,Ma2018GaAs}. In wide-gap GaN, room-temperature hole mobilities are limited to merely tens of cm$^2$/Vs~\cite{Arakawa2016HighHole,Horita2017HallEffect}.
Recent calculations and experiments have established cubic BAs as a rare high-mobility ambipolar semiconductor~\cite{Liu2018SimultaneouslyHigh,Yue2022HighAmbipolar,Shin2022BAsAmbipolarMobility}, but its phonon-limited hole mobility reaches only about 2000~cm$^2$/Vs at 300~K~\cite{Liu2018SimultaneouslyHigh}.
Even in diamond, the leading bulk benchmark for hole transport, the measured room-temperature hole mobility does not exceed 3800~cm$^2$/Vs~\cite{Isberg2002HighCarrier}.
The challenge is therefore to discover bulk semiconductors with substantially higher hole mobility and to identify mechanisms that can guide the design of high-mobility materials. \\
%
%
\indent
Previous efforts to improve hole mobility have generally followed two routes.
One engineers lighter valence bands,  for example through strain, crystal-field control, or chemical design that pushes heavy bands away from the valence-band maximum (VBM)~\cite{Williamson2017EngineeringValenceBand,Ponce2019Route,Chen2026HighHoleMobility}; 
the other weakens long-range Fr\"ohlich electron-phonon (e-ph) coupling by favoring stiff covalent lattices with weak polar interactions~\cite{Liu2018SimultaneouslyHigh,Zhang2023TwoDimensionalSemiconductors}.
Both can lower effective masses or suppress small-$\bm q$ scattering, 
but neither addresses intervalley scattering when several symmetry-related valleys lie near the band edge.
A distinct mechanism emerged in monolayer transition-metal dichalcogenides, where broken inversion symmetry and spin-orbit 
coupling (SOC) lock opposite spins to time-reversed valleys~\cite{Xiao2012CoupledSpin};
because phonon perturbations are largely spin independent, this spin-valley locking (SVL) suppresses intervalley scattering and enhances hole mobility.
It has so far been demonstrated in non-centrosymmetric two-dimensional (2D) materials~\cite{Ciccarino2018DynamicsSpin,Ha2025UltrahighHole}, since it requires a valley spin splitting enabled by inversion symmetry breaking.
Inversion-symmetric crystals composed of locally non-centrosymmetric sectors can nevertheless host hidden spin polarization~\cite{Zhang2014HiddenSpin,Guan2022ProgressHiddenSpin,Xiong2026ApparentHidden}, with opposite spin textures residing on inversion-partner sectors and compensating exactly in the total band structure.
Such textures have been observed spectroscopically~\cite{Riley2014DirectObservation,Huang2020HiddenSpin,Zhang2021ObservationSpinMomentumLayer, Arnoldi2024RevealingHidden,Okuda2026Hidden,Chakraborty2026HiddenSpinValley}, but whether they can suppress carrier scattering and enhance charge transport in a centrosymmetric bulk semiconductor remains unresolved. \\
%
%
\begin{figure*}[!htbp]
\centering
\includegraphics[width=2.05\columnwidth]{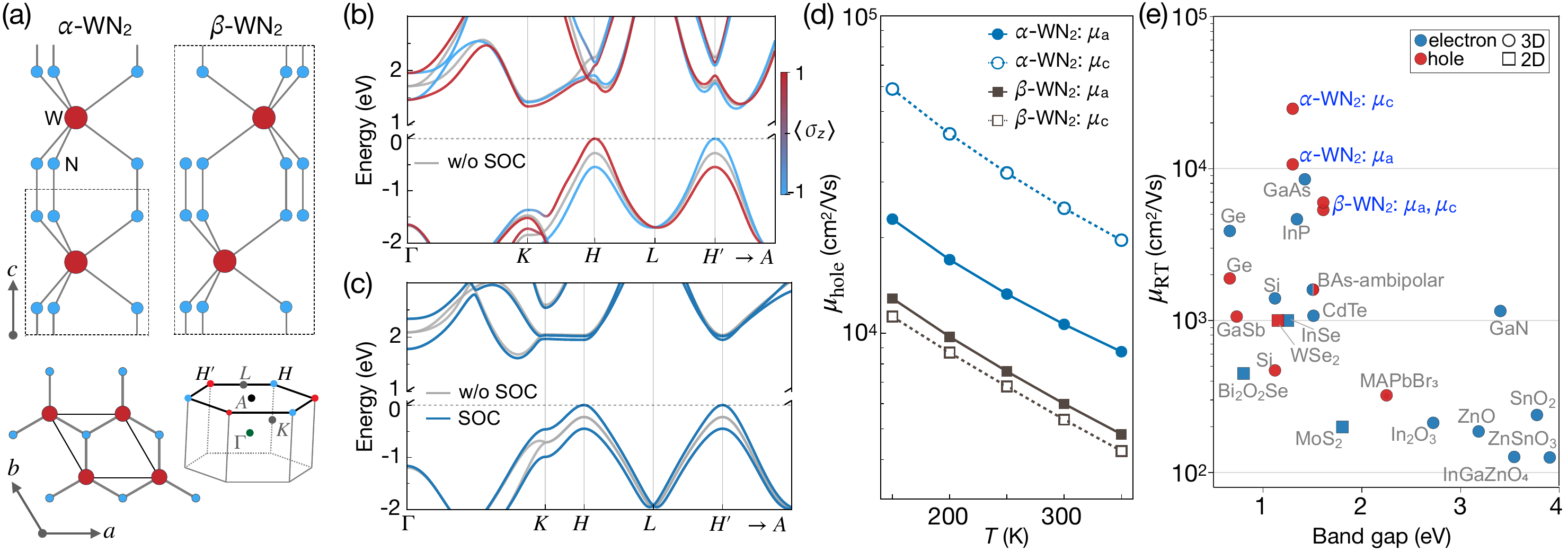}
\caption{\label{fig:fig1}%
(a)~Crystal structures of $\alpha$- and $\beta$-WN$_2$, and the corresponding Brillouin zone. Dashed boxes mark the primitive cells.
(b)~Band structure of $\alpha$-WN$_2$ without SOC (grey) and with SOC
(colored by $\langle\sigma_z\rangle$). Red and blue denote spin-up and spin-down states, respectively.
(c)~Band structure of $\beta$-WN$_2$ without SOC (grey) and with SOC (blue).
(d)~Calculated temperature-dependent hole mobilities of both phases. Filled and
open symbols denote in-plane ($\mu_{a}$) and out-of-plane ($\mu_{c}$) mobilities, respectively.
(e)~Room-temperature carrier mobility as a function of band gap, comparing the calculated results of WN$_2$ with experimental data of representative bulk and 2D 
semiconductors~\cite{Wolfe1970ElectronMobility,Yue2022HighAmbipolar, Galazka2021OxideHallMobility,
Prince1953GermaniumMobility, *Segall1963CdTeElectrical, *Jacoboni1977SiliconTransport, Engh1981InPCrystal, 
Gu2016GaNElectrical, Bandurin2017InSeMobility, Kimura2021MAPbBr3Mobility,  Zhao2024GaSbMobility,
Wu2017Bi2O2SeMobility,Radisavljevic2011MoS2Transistors,Pack2024WSe2ChargeTransferContacts}.}
\end{figure*}
%
%
\indent
In this Letter, we show that hidden spin polarization can suppress intervalley scattering in a centrosymmetric bulk 
material as effectively as net SVL does in a non-centrosymmetric one. 
We demonstrate this in two hexagonal phases of bulk WN$_2$, the non-centrosymmetric 
$\alpha$-phase and centrosymmetric $\beta$-phase, both sizable-gap semiconductors.
Our \textit{ab initio} calculations predict ultrahigh room-temperature hole mobilities exceeding 
10000~cm$^2$/Vs in $\alpha$-WN$_2$ and about 6000~cm$^2$/Vs in $\beta$-WN$_2$.
In the $\alpha$-phase, SOC-induced large valley spin splitting produces net SVL at the valence-band edge, whereas in the $\beta$-phase, hidden Zeeman-type spin polarization yields a compensated, sector-resolved spin texture that reverses between time-reversed valleys.
Both the net and hidden forms of SVL nearly eliminate intervalley e-ph scattering, raising the hole mobility by about an order of magnitude in both phases.
The remaining intravalley scattering is intrinsically weak, owing to the stiff W--N/N--N covalent network, 
and is governed mainly by quadrupole e-ph interactions.
These results establish hidden spin polarization as an active transport-protection mechanism and 
identify locally spin-locked valleys as a design principle for high-mobility materials.

%% file: result-1.tex


Figure~\ref{fig:fig1}(a) shows the crystal structures of the two hexagonal WN$_2$ phases. 
The non-centrosymmetric $\alpha$-phase adopts the WC-type $P\bar{6}\text{m}2$ structure, while the centrosymmetric $\beta$-phase adopts the NiAs-type $P6_3/\text{mmc}$ structure. 
In both phases, the short axial N--N dumbbells have strong covalent single-bond character and can be formally described as $[\text{N}_2]^{4-}$ units, which connect the W--N framework along the $c$ direction and form a three-dimensional covalent network rather than weakly coupled layers.
Both phases are predicted to be dynamically stable at ambient pressure and accessible under high-pressure synthesis conditions, with the $\beta$-phase slightly lower in formation enthalpy~\cite{Wang2009Ultraincompressible}.

We compute the electronic structures, phonon dispersions, and e-ph perturbation potentials with the PBEsol exchange-correlation functional in~\textsc{Quantum Espresso}~\cite{Perdew2008Restoring, Giannozzi2017Advanced}.
Band structures are additionally obtained with the HSE06 hybrid functional in~\textsc{VASP}~\cite{Krukau2006Influence,Kresse1996Efficient}, and the HSE06-corrected band energies enter the scattering-rate and transport calculations~\cite{Lee2018naphthalene,Abramovitch2023,Hu2026Phonon}.
The e-ph matrix elements are evaluated on coarse $\bm{k}$- and $\bm{q}$-grids and Wannier-interpolated to ultrafine meshes using~\textsc{Perturbo}~\cite{Zhou2021Perturbo}, including the long-range dipole and quadrupole contributions~\cite{Jhalani2020PiezoelectricElectron,Park2020LongRangeQuadrupole,Brunin2020PRL,Brunin2020PRB,Sjakste2015,Verdi2015}. We carefully validate the interpolated values against direct density functional perturbation theory calculations~\cite{Baroni2001RMP}. 
Phonon-limited hole mobilities are then obtained by iteratively solving the linearized Boltzmann transport equation in \textsc{Perturbo}~\cite{Zhou2021Perturbo}.
Further details are given in the Supplemental Material (SM)~\cite{SupplementalMaterial}.

Figures~\ref{fig:fig1}(b) and~\ref{fig:fig1}(c) compare band structures of the two phases with and without SOC from HSE06 calculations.
Both $\alpha$- and $\beta$-WN$_2$ are indirect-gap semiconductors with VBM at the time-reversal-related $H$ and $H^{\prime}$ of the Brillouin zone (BZ). 
These valley-edge states appear already without SOC and derive mainly from hybridization between W-$d_{xy}/d_{x^2-y^2}$ and N-$p_x/p_y$ orbitals. 
Because the $\alpha$-phase lacks inversion symmetry, SOC induces a large valley spin splitting of  $\sim$0.54~eV, with opposite out-of-plane spin polarizations 
at $H$ and $H^{\prime}$ [see Fig.~\ref{fig:fig1}(b)]. 
This net SVL and the dominant $d_{xy}/d_{x^2-y^2}$ orbital character closely resemble those of monolayer transition-metal dichalcogenides~\cite{Xiao2012CoupledSpin,Liu2013ThreeBand}, but occur here in a bulk crystal at valleys on the $k_z$~=~$\pi$ plane.
In the $\beta$-phase, by contrast, inversion and time-reversal symmetries enforce twofold band degeneracy throughout the BZ; 
SOC nonetheless separates the upper and lower valence doublets at $H/H^{\prime}$ by $\sim$0.45~eV [Fig.~\ref{fig:fig1}(c)], each with zero net spin polarization.

These distinct valley spin structures lead to a common transport outcome, with both phases exhibiting exceptionally high phonon-limited hole mobilities.
Figure~\ref{fig:fig1}(d) shows the calculated in-plane and out-of-plane mobilities from 150 to 350~K, all of which decrease approximately as $T^{-1.2}$ over this temperature range.
At 300~K, the $\alpha$-phase reaches $\mu_a$~$\approx$~10600 and $\mu_c$~$\approx$~24700~cm$^2$/Vs, while the nearly isotropic $\beta$-phase reaches $\mu_a$~$\approx$~6000 and $\mu_c$~$\approx$~5300~cm$^2$/Vs. 
The mobility anisotropy roughly follows the hole effective masses, which are 0.34$m_0$ and 0.25$m_0$ along $a$ and $c$ directions for $\alpha$-WN$_{2}$, and 0.35$m_0$ and 0.52$m_0$ for $\beta$-WN$_{2}$.
Unlike monolayer systems where SVL is established~\cite{Xiao2012CoupledSpin,Ciccarino2018DynamicsSpin,Ha2025UltrahighHole}, WN$_2$ sustains high hole mobility along both principal directions, a genuinely bulk effect.

\begin{figure}[!htbp]
  \centering
  \includegraphics[width=\columnwidth]{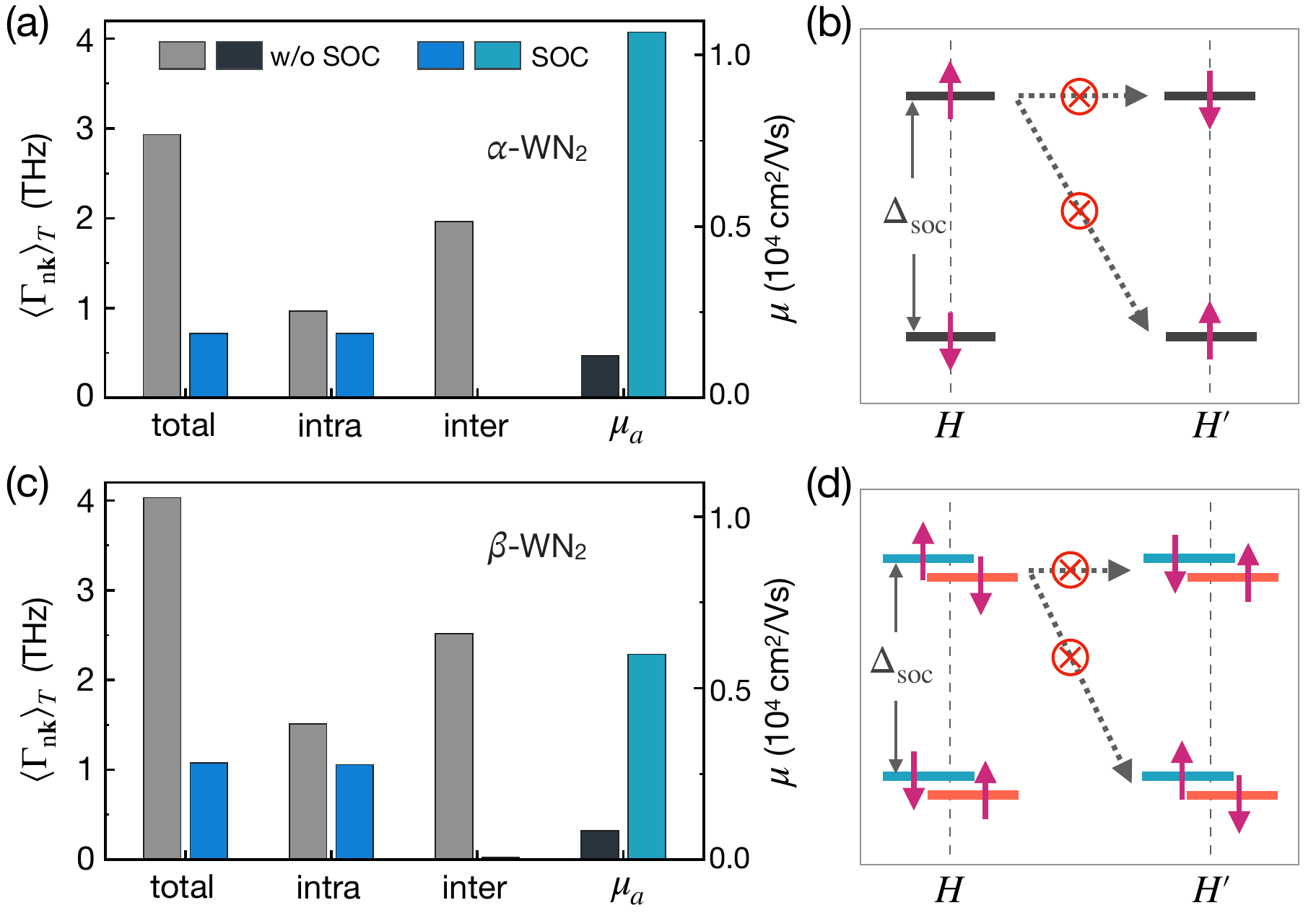}
  \caption{\label{fig:fig2}%
  (a,c)~Thermally averaged total, intravalley, and intervalley scattering rates, 
  together with the hole mobility $\mu_a$, calculated at 300~K without and with SOC, for
  $\alpha$-WN$_2$ in (a) and $\beta$-WN$_2$ in (c).
  (b,d)~Schematics of net and hidden SVL at $H$ and $H^{\prime}$ in
  $\alpha$- and $\beta$-WN$_2$, respectively. Magenta arrows indicate the out-of-plane spin
  direction, and crossed arrows mark suppressed intervalley transitions. In (d), cyan and
  coral lines denote the dominant A and B sector characters, respectively.}
\end{figure}

The ultrahigh hole mobility and sizable band gap place WN$_2$ in an unusual region of the semiconductor mobility--gap landscape.
Figure~\ref{fig:fig1}(e) compares the predicted WN$_2$ hole mobilities with measured mobilities of representative bulk and 2D materials at room temperature.
With gaps of 1.3 and 1.6~eV from HSE06 calculations, $\alpha$- and $\beta$-WN$_2$ exhibit hole mobilities in the range usually associated with high electron mobility.
To our knowledge, these are the highest room-temperature hole mobilities predicted from first-principles e-ph calculations for a sizable-gap bulk semiconductor. 
In particular, both principal-axis hole mobilities of $\alpha$-WN$_2$ exceed the electron mobility of GaAs, even though GaAs has a much smaller electron effective mass of $\sim$0.06~$m_0$~\cite{Zhou2016AbInitio,Liu2017GaAs,Ma2018GaAs}. Therefore, light carrier masses alone cannot explain these ultrahigh mobilities; weak e-ph scattering must instead be responsible.
The comparably high mobility of the spin-degenerate $\beta$-phase further indicates that net SVL is not necessary to access this regime.

%% file: result-2.tex


\begin{figure}[!htbp]
  \centering
  \includegraphics[width=\columnwidth]{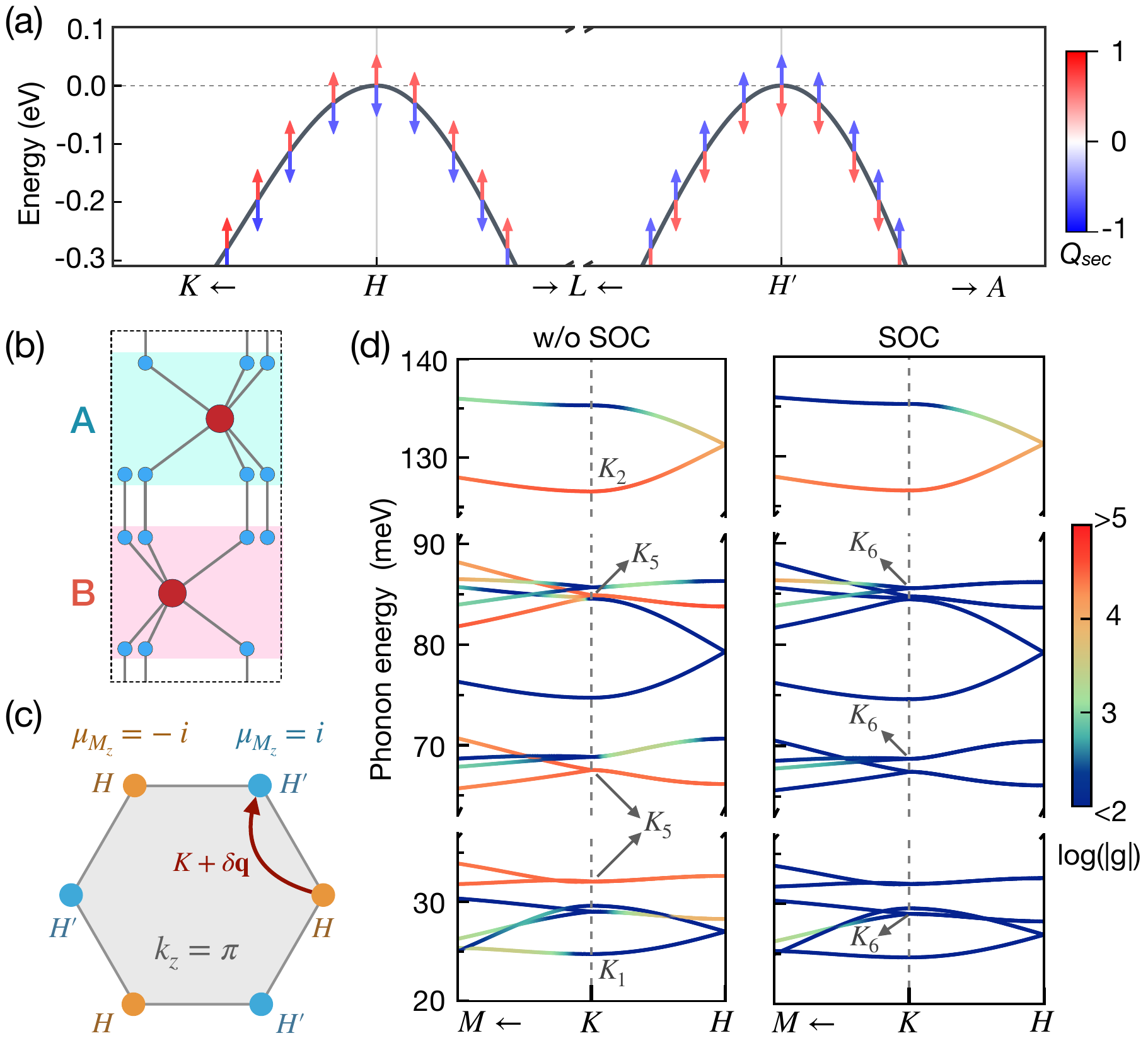}
  \caption{\label{fig:fig3}%
  (a)~Top valence bands around $H$ and $H^{\prime}$ in $\beta$-WN$_2$.
  For each doublet in the $\sigma_z$ gauge, arrow direction (length) encodes the sign (magnitude) 
  of $\langle\sigma_z\rangle$, while color gives the sector polarization
  $Q_{\rm sec}$ defined in Eq.~(\ref{eq:qsec}).
  (b)~The unit cell shows the inversion-partner sectors $A$ and $B$, 
  which are connected by N--N bonds.
  (c)~The $k_z$~=~$\pi$ plane of the first BZ, where $H$ and $H^{\prime}$ have eigenvalues
  $\mu_{M_z}$~=~$-i$ and $+i$, respectively, under the shifted 
  mirror $M_z$~=~$\{m_z|00\tfrac{1}{2}\}$. 
  The arrow denotes the intervalley phonon wave vector $\bm q$~=~$\bm{K}+\delta\bm q$.
  (d)~Phonon dispersions colored by the intervalley e-ph coupling strength $|g_\nu(H,\bm q)|$~\cite{SupplementalMaterial} in log-scale, without and with SOC.
  Selected phonon irreducible representations at $K$ are labeled.}
\end{figure}

\textit{Net and hidden SVL}---
We now turn to the microscopic origin of the weak e-ph scattering, starting with the net SVL mechanism in the non-centrosymmetric $\alpha$-phase. 
As shown in Fig.~\ref{fig:fig1}(b) and sketched in Fig.~\ref{fig:fig2}(b), the upper valley-edge states at $H$ and $H^{\prime}$ carry opposite spin polarizations.
Because e-ph interactions conserve spin to leading order, the direct transition between the two valleys is spin-flip-like and effectively forbidden.
A spin-conserving transition could instead reach the lower SOC branch at the opposite valley, but its separation from the valence edge, $\Delta_{\text{SOC}}$~$\sim$~0.54~eV, well exceeds the highest phonon energy of $\sim$0.14~eV~\cite{SupplementalMaterial}.
Spin and energy conservation therefore act together to suppress intervalley scattering.

To verify this mechanism at the transport level, Fig.~\ref{fig:fig2}(a) compares the thermally averaged e-ph scattering rates with and without SOC, resolved into intravalley and intervalley contributions, together with the corresponding hole mobilities. 
Intravalley processes connect states within the same $H$ or $H^{\prime}$ valley through small-$\bm q$ phonons, whereas intervalley processes transfer holes between valleys through phonons with wave vectors near $K$ [Fig.~\ref{fig:fig3}(c)].
The thermally averaged rates are obtained by averaging each state-resolved scattering rate over the thermally active hole states at 300~K and a fixed hole concentration of $10^{18}$~cm$^{-3}$, see Eq.~(S10) in SM~\cite{SupplementalMaterial}.
Note that this average excludes the band-velocity factor, separating changes in the collision rates from the velocity weighting that enters the mobility. 
Without SOC, intervalley processes account for about two thirds of the total rate.
SOC nearly removes the intervalley processes while reducing the intravalley rate by $\sim$30\%, and enhances the hole mobility by more than a factor of eight. The mobility enhancement exceeds the reduction in total scattering rate, reflecting the velocity weighting and momentum-relaxation factors absent from the thermally averaged rate~\cite{SupplementalMaterial}.
The $\alpha$-phase thus establishes a direct link between net SVL and the high mobility.

Figure~\ref{fig:fig2}(c) shows a similar response to SOC in the $\beta$-phase. 
SOC again nearly eliminates the dominant intervalley e-ph processes, reducing the thermally averaged total rate to about one quarter of its no-SOC value, and enhancing the hole mobility by more than a factor of seven.
This channel-selective response cannot arise from conventional net SVL, since inversion and time-reversal symmetries enforce twofold band degeneracy, so both spin states are available at the same energy within each valley doublet.
It instead points to a compensated, sector-resolved form of SVL.

This hidden SVL is shown schematically in Fig.~\ref{fig:fig2}(d). Its microscopic origin lies in the valence states before SOC. 
The no-SOC $H$-valley edge already contains two degenerate conjugate crystal-orbitals that arise from hybridization between the chiral W-$d_{\pm2}$ and N-$p_{\pm}$ states, where
\[
d_{\pm2}=\tfrac{1}{\sqrt{2}}\left(d_{x^2-y^2}\pm i d_{xy}\right),
\qquad
p_{\pm}=\tfrac{1}{\sqrt{2}}\left(p_x\pm i p_y\right).
\]
Because inversion ($P$) and time-reversal ($\mathcal{T}$) each exchange $H$ and $H^{\prime}$, their product leaves $H$ invariant. The orbital basis can therefore be chosen such that $\ket{u_-}=P\mathcal{T}\ket{u_+}$, with the two states written schematically as
\[
\ket{u_+}\sim\ket{W^A_{d_{+2}}}+\ket{N^A_{p_+}},
\qquad
\ket{u_-}\sim\ket{W^B_{d_{-2}}}+\ket{N^B_{p_-}}.
\]
These two branches carry opposite orbital chirality and complementary character on the inversion-partner $A$ and $B$ sectors shown in Fig.~\ref{fig:fig3}(b). 
Note that the A and B sectors are connected by N--N bonds within the bulk covalent network, rather than separate layers. 
Including spin gives the four-state manifold $\{\ket{u_+},\ket{u_-}\}\otimes\{\ket{\uparrow},\ket{\downarrow}\}$. 
The leading Ising-type SOC component, $\lambda_{\text{soc}} L_z S_z$, splits this manifold into an upper pair dominated by $\ket{u_+\uparrow}$ and $\ket{u_-\downarrow}$, and a lower pair dominated by $\ket{u_+\downarrow}$ and $\ket{u_-\uparrow}$.
SOC therefore correlates spin with the pre-existing orbital-sector character. 

Within this sector-spin manifold, the effective valley-edge Hamiltonian takes the form
\begin{equation}
H_{\eta}\simeq E_{0}I+\Delta_{0}\eta\tau_zs_z,
\label{eq:hidden-svl}
\end{equation}
where $\eta$~=~$\pm1$ labels $H/H^{\prime}$, $s_z$~=~$\pm1$ denotes the out-of-plane spin, and $\tau_z$~=~$\pm1$ labels branches with dominant $A/B$ sector character.
Inversion $P$ reverses both $\eta$ and $\tau_z$ while preserving spin $s_z$, whereas time reversal $\mathcal T$ reverses $\eta$ and $s_z$ while preserving the orbital-sector $\tau_z$.
Their product $P\mathcal T$ leaves $\eta$ unchanged while reversing both $\tau_z$ and $s_z$, therefore it pairs opposite-sector, opposite-spin states within the same valley and keeps them degenerate~[see Fig.~\ref{fig:fig2}(d)]. 

This compensated, sector-resolved spin texture realizes the hidden Zeeman-type spin polarization established for centrosymmetric crystals with twofold screw-rotational $S_{2z}$ symmetry~\cite{Guan2023HiddenZeeman,Wu2018nodal}, a symmetry that $\beta$-WN$_2$ possesses.
Combining $S_{2z}$ with inversion gives the shifted mirror $M_z$~=~$S_{2z}P$ ~=~$\{m_z|00\tfrac12\}$, which leaves the $k_z$~=~$\pi$ plane invariant and constrains the states at $H$ and $H^{\prime}$.
On the $k_z$~=~$\pi$ plane, $M_z$ reverses the in-plane spin components while preserving $\braket{\sigma_z}$, thus enforces $\braket{\sigma_x}$~=~$\braket{\sigma_y}$~=~0, which makes the eigenstates of $\sigma_z$ the natural spin gauge. We take the eigenstates of $\sigma_z$ as the doublet states and define their sector polarizations as~\cite{SupplementalMaterial}
\begin{equation}
\sigma_z|u_\lambda\rangle=s_\lambda|u_\lambda\rangle,
\qquad
Q_{\rm sec}^{(\lambda)}
=\langle u_\lambda|P_A-P_B|u_\lambda\rangle,
\label{eq:qsec}
\end{equation}
where $P_A$ and $P_B$ project onto the two inversion-partner structural sectors of the unit cell~[Fig.~\ref{fig:fig3}(b)]. 
We compute the spin and sector polarizations of each state within the upper valence doublets~[Fig.~\ref{fig:fig3}(a)]. 
The two $P\mathcal T$ partners at each valley carry opposite spins and opposite sector polarizations, and this spin-sector association reverses between $H$ and $H^{\prime}$ valley while the net spin polarization of each doublet vanishes.
The calculated doublets thus reproduce the sector-resolved spin structure predicted by Eq.~\eqref{eq:hidden-svl} and sketched in Fig.~\ref{fig:fig2}(d).
Although Fig.~\ref{fig:fig3}(a) is plotted in a particular spin gauge, the spin-sector association itself is gauge invariant within the degenerate doublet~\cite{SupplementalMaterial}.

The sector-resolved spin structure provides an intuitive picture of the intervalley scattering constraint.
A same-sector transition between $H$ and $H^{\prime}$ is spin-flip-like, whereas a spin-conserving transition must connect opposite sectors.
The rigorous selection rule follows from the shifted mirror $M_z$ symmetry discussed above.
For spinful states on the $k_z$~=~$\pi$ plane, $M_z^2$~=~$-1$ restricts the shifted-mirror eigenvalues to $\pm i$, while the projective relation $M_zP\mathcal T$~=~$-P\mathcal TM_z$ ensures that the two $P\mathcal T$ partners within each doublet share the same eigenvalue.
Time reversal $\mathcal T$ maps $H$ to $H^{\prime}$ and complex-conjugates this eigenvalue.
Thus, the top valence doublets of the $H$ and $H^{\prime}$ valley carry opposite scalar mirror eigenvalues of $-i$ and $+i$, respectively [Fig.~\ref{fig:fig3}(c)]. 
For a phonon branch $\nu$ with mirror parity $\xi_\nu$, the intervalley e-ph vertex obeys $G^{fi}_\nu$~=~$\xi_\nu\mu_f^*\mu_iG^{fi}_\nu$, where $\mu_i$ ($\mu_f$) is the mirror eigenvalue of the initial (final) state. Because $\mu_f^*\mu_i$~=~$-1$ for $H$~$\rightarrow$~$H^{\prime}$,
the vertex vanishes for mirror-even ($\xi_\nu$~=~1) branches provided both electronic states lie on the $k_z$~=~$\pi$ plane.
%
The mirror eigenvalues are symmetry labels of the full doublet, so this constraint is independent of the gauge choice.
Mirror-odd processes remain allowed, therefore the shifted mirror symmetry selects which intervalley channels survive rather than forbidding them all. 
 
Figure~\ref{fig:fig3}(d) confirms this constraint in the mode-resolved intervalley e-ph coupling strength $\lvert g_\nu \rvert$~\cite{SupplementalMaterial}.
Intervalley scattering involves phonons with wave vector $\bm q$~=~$\bm{K}+\delta\bm q$, where $\delta\bm q$ is small [Fig.~\ref{fig:fig3}(c)].  
At $\delta\bm q$=$\bm 0$, the phonon symmetries compatible with the transition change from $K_2\oplus K_3\oplus K_5$ without SOC to $2K_2\oplus2K_3$ with SOC~\cite{SupplementalMaterial}. 
Without SOC, the strongest couplings are concentrated in several low-energy $K_5$ branches and a $K_2$ branch near 126~meV.
Including SOC, the shifted-mirror selection rule excludes the mirror-even $K_5$ branches, and their intervalley coupling vanishes.
The mirror-odd $K_2$ branch survives but contributes negligibly to intervalley scattering, since its energy, about five times $k_{B}T$ at 300~K, strongly limits both its thermal occupation and the phase space for emission.

This protection is not confined to exact $\bm K$. 
For an in-plane deviation $\delta\bm q_\parallel$, $M_z$ remains an exact symmetry and mirror-even intervalley coupling still vanishes.
An out-of-plane component $\delta q_\perp$ breaks $M_z$ and allows mirror-even amplitudes to reappear at linear order.
However, since the rate depends on $\lvert g \rvert^2$, these amplitudes contribute only as $\delta q_\perp^2$, and their effect on the intervalley rate is negligible~\cite{SupplementalMaterial}. 
The removal of the dominant zeroth-order channels thus explains the nearly vanishing intervalley rate in Fig.~\ref{fig:fig2}(c). 
We note that the in-plane protection follows from the shifted-mirror symmetry rather than a material-specific cancellation, so the mechanism should extend to other centrosymmetric crystals with hidden Zeeman-type spin polarization.

%% file: result-3.tex


\begin{figure}[!htbp]
  \centering
  \includegraphics[width=\columnwidth]{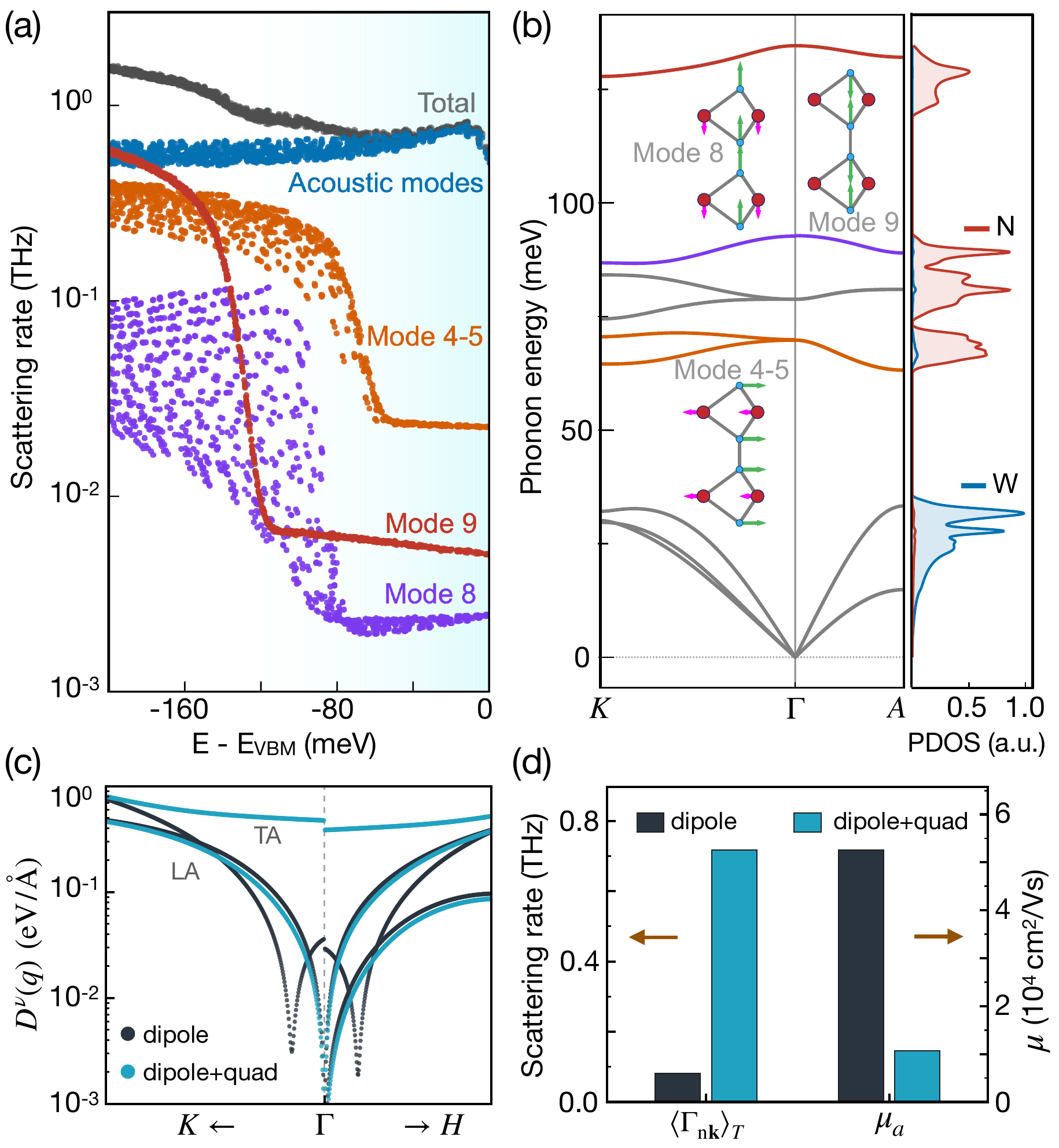}
  \caption{\label{fig:fig4}%
  Intravalley scattering and the quadrupole interactions in $\alpha$-WN$_2$.
  (a)~Total and mode-resolved e-ph scattering rates at 300~K, as a function of valence band energy.
  (b)~Phonon dispersion, W- and N-projected phonon densities of states, and displacement patterns of the selected modes.
  (c)~Acoustic deformation potentials $D^\nu(H,\bm q)$~\cite{SupplementalMaterial}, and (d)~thermally averaged total scattering rate and 
  in-plane hole mobility at 300~K, calculated without and with quadrupole interactions.}
\end{figure}

\textit{Intravalley scattering}---
Once intervalley scattering is suppressed, intravalley processes become the remaining transport bottleneck.
To determine why the residual intravalley scattering stays weak, we resolve the near-edge scattering rate of $\alpha$-WN$_2$ into phonon-mode contributions in Fig.~\ref{fig:fig4}(a).
Within the thermally active energy range near the VBM, the three acoustic branches together account for nearly the entire rate, while the representative optical modes contribute at least an order of magnitude less, including the polar-optical phonons.
The phonon dispersion and projected density of states in Fig.~\ref{fig:fig4}(b) show that the low-energy acoustic phonons are well separated from the higher-frequency optical modes dominated by N motion, consistent with the stiff W-N/N-N covalent network.
These high optical frequencies keep the corresponding modes weakly populated near room temperature.
The same covalent bonding also suppresses the Fr\"ohlich coupling. Because the bonding electrons follow the displaced ion closely, the electronic and ionic contributions to the polarization nearly cancel, yielding small Born effective charges below $0.44\,e$~\cite{SupplementalMaterial}. 
The high optical frequencies and weak Fr\"ohlich coupling limit the optical contribution to the scattering rate, leaving acoustic phonons as the dominant scattering mechanism.

The residual acoustic coupling cannot be described accurately without the dynamical quadrupole interactions~\cite{Jhalani2020PiezoelectricElectron,Park2020LongRangeQuadrupole,Brunin2020PRL,Brunin2020PRB}.
To quantify its contribution, we compare the acoustic coupling with and without the quadrupole term in Fig.~\ref{fig:fig4}(c) through the e-ph deformation potential~[see Eq.~(S2) in SM~\cite{SupplementalMaterial}].
Including the quadrupole enhances the long-wavelength acoustic coupling and substantially raises the thermally averaged scattering rate [Fig.~\ref{fig:fig4}(d)].
The in-plane mobility accordingly decreases by a factor of five, from about 53500 to 10600~cm$^2$/Vs.
Although Fig.~\ref{fig:fig4} uses the $\alpha$-phase as a representative example, the same quadrupole interactions applies to the $\beta$-phase and is included in its reported mobility.
In both phases, the covalent lattice keeps the intravalley acoustic scattering weak, with the quadrupole interactions as its dominant contribution, and even with these interactions the rates remain low enough to sustain room-temperature hole mobility in $\alpha$-WN$_2$ above $10^{4}$~cm$^2$/Vs.

%% file: conclusion.tex

In summary, we have shown that both net and hidden SVL enable ultrahigh room-temperature hole mobility in the two hexagonal phases of bulk WN$_2$.
In each phase, the valley spin texture nearly eliminates intervalley e-ph scattering, while the stiff covalent network leaves only weak intravalley scattering via acoustic phonons dominated by quadrupole interactions.
Hidden spin polarization has been explored mainly as a band-structure or optical phenomenon.
Our results demonstrate that it also controls the microscopic scattering processes responsible for charge dissipation. Global spin compensation does not erase the role of local spin structure in carrier dynamics.
The broad classes of hidden spin physics identified recently open a much larger materials space in which to exploit similar mechanisms~\cite{Xiong2026ApparentHidden,Zhang2025CrystalSymmetryPaired}. Transport-active hidden spin polarization may offer a general route to high hole mobility in centrosymmetric semiconductors.

%% file: manuscript.bbl
\begin{thebibliography}{71}%
\makeatletter
\providecommand \@ifxundefined [1]{%
 \@ifx{#1\undefined}
}%
\providecommand \@ifnum [1]{%
 \ifnum #1\expandafter \@firstoftwo
 \else \expandafter \@secondoftwo
 \fi
}%
\providecommand \@ifx [1]{%
 \ifx #1\expandafter \@firstoftwo
 \else \expandafter \@secondoftwo
 \fi
}%
\providecommand \natexlab [1]{#1}%
\providecommand \enquote  [1]{``#1''}%
\providecommand \bibnamefont  [1]{#1}%
\providecommand \bibfnamefont [1]{#1}%
\providecommand \citenamefont [1]{#1}%
\providecommand \href@noop [0]{\@secondoftwo}%
\providecommand \href [0]{\begingroup \@sanitize@url \@href}%
\providecommand \@href[1]{\@@startlink{#1}\@@href}%
\providecommand \@@href[1]{\endgroup#1\@@endlink}%
\providecommand \@sanitize@url [0]{\catcode `\\12\catcode `\$12\catcode
  `\&12\catcode `\#12\catcode `\^12\catcode `\_12\catcode `\%12\relax}%
\providecommand \@@startlink[1]{}%
\providecommand \@@endlink[0]{}%
\providecommand \url  [0]{\begingroup\@sanitize@url \@url }%
\providecommand \@url [1]{\endgroup\@href {#1}{\urlprefix }}%
\providecommand \urlprefix  [0]{URL }%
\providecommand \Eprint [0]{\href }%
\providecommand \doibase [0]{https://doi.org/}%
\providecommand \selectlanguage [0]{\@gobble}%
\providecommand \bibinfo  [0]{\@secondoftwo}%
\providecommand \bibfield  [0]{\@secondoftwo}%
\providecommand \translation [1]{[#1]}%
\providecommand \BibitemOpen [0]{}%
\providecommand \bibitemStop [0]{}%
\providecommand \bibitemNoStop [0]{.\EOS\space}%
\providecommand \EOS [0]{\spacefactor3000\relax}%
\providecommand \BibitemShut  [1]{\csname bibitem#1\endcsname}%
\let\auto@bib@innerbib\@empty
\bibitem [{\citenamefont {del
  Alamo}(2011)}]{DelAlamo2011NanometreScaleElectronics}%
  \BibitemOpen
  \bibfield  {author} {\bibinfo {author} {\bibfnamefont {J.~A.}\ \bibnamefont
  {del Alamo}},\ }\bibfield  {title} {\bibinfo {title} {{Nanometre-scale
  electronics with III--V compound semiconductors}},\ }\href
  {https://doi.org/10.1038/nature10677} {\bibfield  {journal} {\bibinfo
  {journal} {Nature}\ }\textbf {\bibinfo {volume} {479}},\ \bibinfo {pages}
  {317} (\bibinfo {year} {2011})}\BibitemShut {NoStop}%
\bibitem [{\citenamefont {Hautier}\ \emph {et~al.}(2013)\citenamefont
  {Hautier}, \citenamefont {Miglio}, \citenamefont {Ceder}, \citenamefont
  {Rignanese},\ and\ \citenamefont
  {Gonze}}]{Hautier2013IdentificationDesignPrinciples}%
  \BibitemOpen
  \bibfield  {author} {\bibinfo {author} {\bibfnamefont {G.}~\bibnamefont
  {Hautier}}, \bibinfo {author} {\bibfnamefont {A.}~\bibnamefont {Miglio}},
  \bibinfo {author} {\bibfnamefont {G.}~\bibnamefont {Ceder}}, \bibinfo
  {author} {\bibfnamefont {G.-M.}\ \bibnamefont {Rignanese}},\ and\ \bibinfo
  {author} {\bibfnamefont {X.}~\bibnamefont {Gonze}},\ }\bibfield  {title}
  {\bibinfo {title} {{Identification and design principles of low hole
  effective mass p-type transparent conducting oxides}},\ }\href
  {https://doi.org/10.1038/ncomms3292} {\bibfield  {journal} {\bibinfo
  {journal} {Nat. Commun.}\ }\textbf {\bibinfo {volume} {4}},\ \bibinfo {pages}
  {2292} (\bibinfo {year} {2013})}\BibitemShut {NoStop}%
\bibitem [{\citenamefont {Gibbs}\ \emph {et~al.}(2017)\citenamefont {Gibbs},
  \citenamefont {Ricci}, \citenamefont {Li}, \citenamefont {Zhu}, \citenamefont
  {Persson}, \citenamefont {Ceder}, \citenamefont {Hautier}, \citenamefont
  {Jain},\ and\ \citenamefont {Snyder}}]{Gibbs2017EffectiveMass}%
  \BibitemOpen
  \bibfield  {author} {\bibinfo {author} {\bibfnamefont {Z.~M.}\ \bibnamefont
  {Gibbs}}, \bibinfo {author} {\bibfnamefont {F.}~\bibnamefont {Ricci}},
  \bibinfo {author} {\bibfnamefont {G.}~\bibnamefont {Li}}, \bibinfo {author}
  {\bibfnamefont {H.}~\bibnamefont {Zhu}}, \bibinfo {author} {\bibfnamefont
  {K.}~\bibnamefont {Persson}}, \bibinfo {author} {\bibfnamefont
  {G.}~\bibnamefont {Ceder}}, \bibinfo {author} {\bibfnamefont
  {G.}~\bibnamefont {Hautier}}, \bibinfo {author} {\bibfnamefont
  {A.}~\bibnamefont {Jain}},\ and\ \bibinfo {author} {\bibfnamefont {G.~J.}\
  \bibnamefont {Snyder}},\ }\bibfield  {title} {\bibinfo {title} {{Effective
  mass and Fermi surface complexity factor from ab initio band structure
  calculations}},\ }\href {https://doi.org/10.1038/s41524-017-0013-3}
  {\bibfield  {journal} {\bibinfo  {journal} {npj Comput. Mater.}\ }\textbf
  {\bibinfo {volume} {3}},\ \bibinfo {pages} {8} (\bibinfo {year}
  {2017})}\BibitemShut {NoStop}%
\bibitem [{\citenamefont {Blakemore}(1982)}]{Blakemore1982Semiconducting}%
  \BibitemOpen
  \bibfield  {author} {\bibinfo {author} {\bibfnamefont {J.~S.}\ \bibnamefont
  {Blakemore}},\ }\bibfield  {title} {\bibinfo {title} {{Semiconducting and
  other major properties of gallium arsenide}},\ }\href
  {https://doi.org/10.1063/1.331665} {\bibfield  {journal} {\bibinfo  {journal}
  {J. Appl. Phys.}\ }\textbf {\bibinfo {volume} {53}},\ \bibinfo {pages} {R123}
  (\bibinfo {year} {1982})}\BibitemShut {NoStop}%
\bibitem [{\citenamefont {Wolfe}\ \emph {et~al.}(1970)\citenamefont {Wolfe},
  \citenamefont {Stillman},\ and\ \citenamefont
  {Lindley}}]{Wolfe1970ElectronMobility}%
  \BibitemOpen
  \bibfield  {author} {\bibinfo {author} {\bibfnamefont {C.~M.}\ \bibnamefont
  {Wolfe}}, \bibinfo {author} {\bibfnamefont {G.~E.}\ \bibnamefont
  {Stillman}},\ and\ \bibinfo {author} {\bibfnamefont {W.~T.}\ \bibnamefont
  {Lindley}},\ }\bibfield  {title} {\bibinfo {title} {{Electron Mobility in
  High-Purity GaAs}},\ }\href {https://doi.org/10.1063/1.1659368} {\bibfield
  {journal} {\bibinfo  {journal} {J. Appl. Phys.}\ }\textbf {\bibinfo {volume}
  {41}},\ \bibinfo {pages} {3088} (\bibinfo {year} {1970})}\BibitemShut
  {NoStop}%
\bibitem [{\citenamefont {Zhou}\ and\ \citenamefont
  {Bernardi}(2016)}]{Zhou2016AbInitio}%
  \BibitemOpen
  \bibfield  {author} {\bibinfo {author} {\bibfnamefont {J.-J.}\ \bibnamefont
  {Zhou}}\ and\ \bibinfo {author} {\bibfnamefont {M.}~\bibnamefont
  {Bernardi}},\ }\bibfield  {title} {\bibinfo {title} {{Ab initio electron
  mobility and polar phonon scattering in GaAs}},\ }\href
  {https://doi.org/10.1103/PhysRevB.94.201201} {\bibfield  {journal} {\bibinfo
  {journal} {Phys. Rev. B}\ }\textbf {\bibinfo {volume} {94}},\ \bibinfo
  {pages} {201201(R)} (\bibinfo {year} {2016})}\BibitemShut {NoStop}%
\bibitem [{\citenamefont {Liu}\ \emph {et~al.}(2017)\citenamefont {Liu},
  \citenamefont {Zhou}, \citenamefont {Liao}, \citenamefont {Singh},\ and\
  \citenamefont {Chen}}]{Liu2017GaAs}%
  \BibitemOpen
  \bibfield  {author} {\bibinfo {author} {\bibfnamefont {T.-H.}\ \bibnamefont
  {Liu}}, \bibinfo {author} {\bibfnamefont {J.}~\bibnamefont {Zhou}}, \bibinfo
  {author} {\bibfnamefont {B.}~\bibnamefont {Liao}}, \bibinfo {author}
  {\bibfnamefont {D.~J.}\ \bibnamefont {Singh}},\ and\ \bibinfo {author}
  {\bibfnamefont {G.}~\bibnamefont {Chen}},\ }\bibfield  {title} {\bibinfo
  {title} {First-principles mode-by-mode analysis for electron-phonon
  scattering channels and mean free path spectra in {GaAs}},\ }\href
  {https://doi.org/10.1103/PhysRevB.95.075206} {\bibfield  {journal} {\bibinfo
  {journal} {Phys. Rev. B}\ }\textbf {\bibinfo {volume} {95}},\ \bibinfo
  {pages} {075206} (\bibinfo {year} {2017})}\BibitemShut {NoStop}%
\bibitem [{\citenamefont {Ma}\ \emph {et~al.}(2018)\citenamefont {Ma},
  \citenamefont {Nissimagoudar},\ and\ \citenamefont {Li}}]{Ma2018GaAs}%
  \BibitemOpen
  \bibfield  {author} {\bibinfo {author} {\bibfnamefont {J.}~\bibnamefont
  {Ma}}, \bibinfo {author} {\bibfnamefont {A.~S.}\ \bibnamefont
  {Nissimagoudar}},\ and\ \bibinfo {author} {\bibfnamefont {W.}~\bibnamefont
  {Li}},\ }\bibfield  {title} {\bibinfo {title} {First-principles study of
  electron and hole mobilities of {Si} and {GaAs}},\ }\href
  {https://doi.org/10.1103/PhysRevB.97.045201} {\bibfield  {journal} {\bibinfo
  {journal} {Phys. Rev. B}\ }\textbf {\bibinfo {volume} {97}},\ \bibinfo
  {pages} {045201} (\bibinfo {year} {2018})}\BibitemShut {NoStop}%
\bibitem [{\citenamefont {Arakawa}\ \emph {et~al.}(2016)\citenamefont
  {Arakawa}, \citenamefont {Ueno}, \citenamefont {Kobayashi}, \citenamefont
  {Ohta},\ and\ \citenamefont {Fujioka}}]{Arakawa2016HighHole}%
  \BibitemOpen
  \bibfield  {author} {\bibinfo {author} {\bibfnamefont {Y.}~\bibnamefont
  {Arakawa}}, \bibinfo {author} {\bibfnamefont {K.}~\bibnamefont {Ueno}},
  \bibinfo {author} {\bibfnamefont {A.}~\bibnamefont {Kobayashi}}, \bibinfo
  {author} {\bibfnamefont {J.}~\bibnamefont {Ohta}},\ and\ \bibinfo {author}
  {\bibfnamefont {H.}~\bibnamefont {Fujioka}},\ }\bibfield  {title} {\bibinfo
  {title} {{High hole mobility p-type GaN with low residual hydrogen
  concentration prepared by pulsed sputtering}},\ }\href
  {https://doi.org/10.1063/1.4960485} {\bibfield  {journal} {\bibinfo
  {journal} {APL Mater.}\ }\textbf {\bibinfo {volume} {4}},\ \bibinfo {pages}
  {086103} (\bibinfo {year} {2016})}\BibitemShut {NoStop}%
\bibitem [{\citenamefont {Horita}\ \emph {et~al.}(2017)\citenamefont {Horita},
  \citenamefont {Takashima}, \citenamefont {Tanaka}, \citenamefont {Matsuyama},
  \citenamefont {Ueno}, \citenamefont {Edo}, \citenamefont {Takahashi},
  \citenamefont {Shimizu},\ and\ \citenamefont {Suda}}]{Horita2017HallEffect}%
  \BibitemOpen
  \bibfield  {author} {\bibinfo {author} {\bibfnamefont {M.}~\bibnamefont
  {Horita}}, \bibinfo {author} {\bibfnamefont {S.}~\bibnamefont {Takashima}},
  \bibinfo {author} {\bibfnamefont {R.}~\bibnamefont {Tanaka}}, \bibinfo
  {author} {\bibfnamefont {H.}~\bibnamefont {Matsuyama}}, \bibinfo {author}
  {\bibfnamefont {K.}~\bibnamefont {Ueno}}, \bibinfo {author} {\bibfnamefont
  {M.}~\bibnamefont {Edo}}, \bibinfo {author} {\bibfnamefont {T.}~\bibnamefont
  {Takahashi}}, \bibinfo {author} {\bibfnamefont {M.}~\bibnamefont {Shimizu}},\
  and\ \bibinfo {author} {\bibfnamefont {J.}~\bibnamefont {Suda}},\ }\bibfield
  {title} {\bibinfo {title} {{Hall-effect measurements of metalorganic
  vapor-phase epitaxy-grown p-type homoepitaxial GaN layers with various Mg
  concentrations}},\ }\href {https://doi.org/10.7567/JJAP.56.031001} {\bibfield
   {journal} {\bibinfo  {journal} {Jpn. J. Appl. Phys.}\ }\textbf {\bibinfo
  {volume} {56}},\ \bibinfo {pages} {031001} (\bibinfo {year}
  {2017})}\BibitemShut {NoStop}%
\bibitem [{\citenamefont {Liu}\ \emph {et~al.}(2018)\citenamefont {Liu},
  \citenamefont {Song}, \citenamefont {Meroueh}, \citenamefont {Ding},
  \citenamefont {Song}, \citenamefont {Zhou}, \citenamefont {Li},\ and\
  \citenamefont {Chen}}]{Liu2018SimultaneouslyHigh}%
  \BibitemOpen
  \bibfield  {author} {\bibinfo {author} {\bibfnamefont {T.-H.}\ \bibnamefont
  {Liu}}, \bibinfo {author} {\bibfnamefont {B.}~\bibnamefont {Song}}, \bibinfo
  {author} {\bibfnamefont {L.}~\bibnamefont {Meroueh}}, \bibinfo {author}
  {\bibfnamefont {Z.}~\bibnamefont {Ding}}, \bibinfo {author} {\bibfnamefont
  {Q.}~\bibnamefont {Song}}, \bibinfo {author} {\bibfnamefont {J.}~\bibnamefont
  {Zhou}}, \bibinfo {author} {\bibfnamefont {M.}~\bibnamefont {Li}},\ and\
  \bibinfo {author} {\bibfnamefont {G.}~\bibnamefont {Chen}},\ }\bibfield
  {title} {\bibinfo {title} {{Simultaneously high electron and hole mobilities
  in cubic boron-V compounds: BP, BAs, and BSb}},\ }\href
  {https://doi.org/10.1103/PhysRevB.98.081203} {\bibfield  {journal} {\bibinfo
  {journal} {Phys. Rev. B}\ }\textbf {\bibinfo {volume} {98}},\ \bibinfo
  {pages} {081203} (\bibinfo {year} {2018})}\BibitemShut {NoStop}%
\bibitem [{\citenamefont {Yue}\ \emph {et~al.}(2022)\citenamefont {Yue},
  \citenamefont {Tian}, \citenamefont {Sui}, \citenamefont {Mohebinia},
  \citenamefont {Wu}, \citenamefont {Tong}, \citenamefont {Wang}, \citenamefont
  {Wu}, \citenamefont {Zhang}, \citenamefont {Ren} \emph
  {et~al.}}]{Yue2022HighAmbipolar}%
  \BibitemOpen
  \bibfield  {author} {\bibinfo {author} {\bibfnamefont {S.}~\bibnamefont
  {Yue}}, \bibinfo {author} {\bibfnamefont {F.}~\bibnamefont {Tian}}, \bibinfo
  {author} {\bibfnamefont {X.}~\bibnamefont {Sui}}, \bibinfo {author}
  {\bibfnamefont {M.}~\bibnamefont {Mohebinia}}, \bibinfo {author}
  {\bibfnamefont {X.}~\bibnamefont {Wu}}, \bibinfo {author} {\bibfnamefont
  {T.}~\bibnamefont {Tong}}, \bibinfo {author} {\bibfnamefont {Z.}~\bibnamefont
  {Wang}}, \bibinfo {author} {\bibfnamefont {B.}~\bibnamefont {Wu}}, \bibinfo
  {author} {\bibfnamefont {Q.}~\bibnamefont {Zhang}}, \bibinfo {author}
  {\bibfnamefont {Z.}~\bibnamefont {Ren}}, \emph {et~al.},\ }\bibfield  {title}
  {\bibinfo {title} {{High ambipolar mobility in cubic boron arsenide revealed
  by transient reflectivity microscopy}},\ }\href
  {https://doi.org/10.1126/science.abn4727} {\bibfield  {journal} {\bibinfo
  {journal} {Science}\ }\textbf {\bibinfo {volume} {377}},\ \bibinfo {pages}
  {433} (\bibinfo {year} {2022})}\BibitemShut {NoStop}%
\bibitem [{\citenamefont {Shin}\ \emph {et~al.}(2022)\citenamefont {Shin},
  \citenamefont {Gamage}, \citenamefont {Ding}, \citenamefont {Chen},
  \citenamefont {Tian}, \citenamefont {Qian}, \citenamefont {Zhou},
  \citenamefont {Lee}, \citenamefont {Zhou}, \citenamefont {Shi} \emph
  {et~al.}}]{Shin2022BAsAmbipolarMobility}%
  \BibitemOpen
  \bibfield  {author} {\bibinfo {author} {\bibfnamefont {J.}~\bibnamefont
  {Shin}}, \bibinfo {author} {\bibfnamefont {G.~A.}\ \bibnamefont {Gamage}},
  \bibinfo {author} {\bibfnamefont {Z.}~\bibnamefont {Ding}}, \bibinfo {author}
  {\bibfnamefont {K.}~\bibnamefont {Chen}}, \bibinfo {author} {\bibfnamefont
  {F.}~\bibnamefont {Tian}}, \bibinfo {author} {\bibfnamefont {X.}~\bibnamefont
  {Qian}}, \bibinfo {author} {\bibfnamefont {J.}~\bibnamefont {Zhou}}, \bibinfo
  {author} {\bibfnamefont {H.}~\bibnamefont {Lee}}, \bibinfo {author}
  {\bibfnamefont {J.}~\bibnamefont {Zhou}}, \bibinfo {author} {\bibfnamefont
  {L.}~\bibnamefont {Shi}}, \emph {et~al.},\ }\bibfield  {title} {\bibinfo
  {title} {High ambipolar mobility in cubic boron arsenide},\ }\href
  {https://doi.org/10.1126/science.abn4290} {\bibfield  {journal} {\bibinfo
  {journal} {Science}\ }\textbf {\bibinfo {volume} {377}},\ \bibinfo {pages}
  {437} (\bibinfo {year} {2022})}\BibitemShut {NoStop}%
\bibitem [{\citenamefont {Isberg}\ \emph {et~al.}(2002)\citenamefont {Isberg},
  \citenamefont {Hammersberg}, \citenamefont {Johansson}, \citenamefont
  {Wikstr{\"o}m}, \citenamefont {Twitchen}, \citenamefont {Whitehead},
  \citenamefont {Coe},\ and\ \citenamefont
  {Scarsbrook}}]{Isberg2002HighCarrier}%
  \BibitemOpen
  \bibfield  {author} {\bibinfo {author} {\bibfnamefont {J.}~\bibnamefont
  {Isberg}}, \bibinfo {author} {\bibfnamefont {J.}~\bibnamefont {Hammersberg}},
  \bibinfo {author} {\bibfnamefont {E.}~\bibnamefont {Johansson}}, \bibinfo
  {author} {\bibfnamefont {T.}~\bibnamefont {Wikstr{\"o}m}}, \bibinfo {author}
  {\bibfnamefont {D.~J.}\ \bibnamefont {Twitchen}}, \bibinfo {author}
  {\bibfnamefont {A.~J.}\ \bibnamefont {Whitehead}}, \bibinfo {author}
  {\bibfnamefont {S.~E.}\ \bibnamefont {Coe}},\ and\ \bibinfo {author}
  {\bibfnamefont {G.~A.}\ \bibnamefont {Scarsbrook}},\ }\bibfield  {title}
  {\bibinfo {title} {{High Carrier Mobility in Single-Crystal Plasma-Deposited
  Diamond}},\ }\href {https://doi.org/10.1126/science.1074374} {\bibfield
  {journal} {\bibinfo  {journal} {Science}\ }\textbf {\bibinfo {volume}
  {297}},\ \bibinfo {pages} {1670} (\bibinfo {year} {2002})}\BibitemShut
  {NoStop}%
\bibitem [{\citenamefont {Williamson}\ \emph {et~al.}(2017)\citenamefont
  {Williamson}, \citenamefont {Buckeridge}, \citenamefont {Brown},
  \citenamefont {Ansbro}, \citenamefont {Palgrave},\ and\ \citenamefont
  {Scanlon}}]{Williamson2017EngineeringValenceBand}%
  \BibitemOpen
  \bibfield  {author} {\bibinfo {author} {\bibfnamefont {B.~A.~D.}\
  \bibnamefont {Williamson}}, \bibinfo {author} {\bibfnamefont
  {J.}~\bibnamefont {Buckeridge}}, \bibinfo {author} {\bibfnamefont
  {J.}~\bibnamefont {Brown}}, \bibinfo {author} {\bibfnamefont
  {S.}~\bibnamefont {Ansbro}}, \bibinfo {author} {\bibfnamefont {R.~G.}\
  \bibnamefont {Palgrave}},\ and\ \bibinfo {author} {\bibfnamefont {D.~O.}\
  \bibnamefont {Scanlon}},\ }\bibfield  {title} {\bibinfo {title} {{Engineering
  Valence Band Dispersion for High Mobility p-Type Semiconductors}},\ }\href
  {https://doi.org/10.1021/acs.chemmater.6b03306} {\bibfield  {journal}
  {\bibinfo  {journal} {Chem. Mater.}\ }\textbf {\bibinfo {volume} {29}},\
  \bibinfo {pages} {2402} (\bibinfo {year} {2017})}\BibitemShut {NoStop}%
\bibitem [{\citenamefont {Ponc{\'e}}\ \emph {et~al.}(2019)\citenamefont
  {Ponc{\'e}}, \citenamefont {Jena},\ and\ \citenamefont
  {Giustino}}]{Ponce2019Route}%
  \BibitemOpen
  \bibfield  {author} {\bibinfo {author} {\bibfnamefont {S.}~\bibnamefont
  {Ponc{\'e}}}, \bibinfo {author} {\bibfnamefont {D.}~\bibnamefont {Jena}},\
  and\ \bibinfo {author} {\bibfnamefont {F.}~\bibnamefont {Giustino}},\
  }\bibfield  {title} {\bibinfo {title} {{Route to High Hole Mobility in GaN
  via Reversal of Crystal-Field Splitting}},\ }\href
  {https://doi.org/10.1103/PhysRevLett.123.096602} {\bibfield  {journal}
  {\bibinfo  {journal} {Phys. Rev. Lett.}\ }\textbf {\bibinfo {volume} {123}},\
  \bibinfo {pages} {096602} (\bibinfo {year} {2019})}\BibitemShut {NoStop}%
\bibitem [{\citenamefont {Chen}\ \emph {et~al.}(2026)\citenamefont {Chen},
  \citenamefont {Wang}, \citenamefont {Chang}, \citenamefont {Dill},
  \citenamefont {Xing}, \citenamefont {Kioupakis},\ and\ \citenamefont
  {Giustino}}]{Chen2026HighHoleMobility}%
  \BibitemOpen
  \bibfield  {author} {\bibinfo {author} {\bibfnamefont {J.-C.}\ \bibnamefont
  {Chen}}, \bibinfo {author} {\bibfnamefont {A.}~\bibnamefont {Wang}}, \bibinfo
  {author} {\bibfnamefont {C.~F.~C.}\ \bibnamefont {Chang}}, \bibinfo {author}
  {\bibfnamefont {J.~E.}\ \bibnamefont {Dill}}, \bibinfo {author}
  {\bibfnamefont {H.~G.}\ \bibnamefont {Xing}}, \bibinfo {author}
  {\bibfnamefont {E.}~\bibnamefont {Kioupakis}},\ and\ \bibinfo {author}
  {\bibfnamefont {F.}~\bibnamefont {Giustino}},\ }\bibfield  {title} {\bibinfo
  {title} {{High hole mobility in AlN from negative crystal-field splitting}},\
  }\href {https://doi.org/10.1063/5.0307823} {\bibfield  {journal} {\bibinfo
  {journal} {Appl. Phys. Lett.}\ }\textbf {\bibinfo {volume} {128}},\ \bibinfo
  {pages} {012102} (\bibinfo {year} {2026})}\BibitemShut {NoStop}%
\bibitem [{\citenamefont {Zhang}\ \emph {et~al.}(2023)\citenamefont {Zhang},
  \citenamefont {Wang}, \citenamefont {Mishra},\ and\ \citenamefont
  {Liu}}]{Zhang2023TwoDimensionalSemiconductors}%
  \BibitemOpen
  \bibfield  {author} {\bibinfo {author} {\bibfnamefont {C.}~\bibnamefont
  {Zhang}}, \bibinfo {author} {\bibfnamefont {R.}~\bibnamefont {Wang}},
  \bibinfo {author} {\bibfnamefont {H.}~\bibnamefont {Mishra}},\ and\ \bibinfo
  {author} {\bibfnamefont {Y.}~\bibnamefont {Liu}},\ }\bibfield  {title}
  {\bibinfo {title} {{Two-Dimensional Semiconductors with High Intrinsic
  Carrier Mobility at Room Temperature}},\ }\href
  {https://doi.org/10.1103/PhysRevLett.130.087001} {\bibfield  {journal}
  {\bibinfo  {journal} {Phys. Rev. Lett.}\ }\textbf {\bibinfo {volume} {130}},\
  \bibinfo {pages} {087001} (\bibinfo {year} {2023})}\BibitemShut {NoStop}%
\bibitem [{\citenamefont {Xiao}\ \emph {et~al.}(2012)\citenamefont {Xiao},
  \citenamefont {Liu}, \citenamefont {Feng}, \citenamefont {Xu},\ and\
  \citenamefont {Yao}}]{Xiao2012CoupledSpin}%
  \BibitemOpen
  \bibfield  {author} {\bibinfo {author} {\bibfnamefont {D.}~\bibnamefont
  {Xiao}}, \bibinfo {author} {\bibfnamefont {G.-B.}\ \bibnamefont {Liu}},
  \bibinfo {author} {\bibfnamefont {W.}~\bibnamefont {Feng}}, \bibinfo {author}
  {\bibfnamefont {X.}~\bibnamefont {Xu}},\ and\ \bibinfo {author}
  {\bibfnamefont {W.}~\bibnamefont {Yao}},\ }\bibfield  {title} {\bibinfo
  {title} {{Coupled Spin and Valley Physics in Monolayers of MoS$_2$ and Other
  Group-VI Dichalcogenides}},\ }\href
  {https://doi.org/10.1103/PhysRevLett.108.196802} {\bibfield  {journal}
  {\bibinfo  {journal} {Phys. Rev. Lett.}\ }\textbf {\bibinfo {volume} {108}},\
  \bibinfo {pages} {196802} (\bibinfo {year} {2012})}\BibitemShut {NoStop}%
\bibitem [{\citenamefont {Ciccarino}\ \emph {et~al.}(2018)\citenamefont
  {Ciccarino}, \citenamefont {Christensen}, \citenamefont {Sundararaman},\ and\
  \citenamefont {Narang}}]{Ciccarino2018DynamicsSpin}%
  \BibitemOpen
  \bibfield  {author} {\bibinfo {author} {\bibfnamefont {C.~J.}\ \bibnamefont
  {Ciccarino}}, \bibinfo {author} {\bibfnamefont {T.}~\bibnamefont
  {Christensen}}, \bibinfo {author} {\bibfnamefont {R.}~\bibnamefont
  {Sundararaman}},\ and\ \bibinfo {author} {\bibfnamefont {P.}~\bibnamefont
  {Narang}},\ }\bibfield  {title} {\bibinfo {title} {{Dynamics and Spin-Valley
  Locking Effects in Monolayer Transition Metal Dichalcogenides}},\ }\href
  {https://doi.org/10.1021/acs.nanolett.8b02300} {\bibfield  {journal}
  {\bibinfo  {journal} {Nano Lett.}\ }\textbf {\bibinfo {volume} {18}},\
  \bibinfo {pages} {5709} (\bibinfo {year} {2018})}\BibitemShut {NoStop}%
\bibitem [{\citenamefont {Ha}\ \emph {et~al.}(2025)\citenamefont {Ha},
  \citenamefont {Tiwari},\ and\ \citenamefont
  {Giustino}}]{Ha2025UltrahighHole}%
  \BibitemOpen
  \bibfield  {author} {\bibinfo {author} {\bibfnamefont {V.-A.}\ \bibnamefont
  {Ha}}, \bibinfo {author} {\bibfnamefont {S.}~\bibnamefont {Tiwari}},\ and\
  \bibinfo {author} {\bibfnamefont {F.}~\bibnamefont {Giustino}},\ }\bibfield
  {title} {\bibinfo {title} {{Ultrahigh Hole Mobility in Monolayer WSe$_2$
  Enabled by Spin-Orbit Suppression of Intervalley Scattering}},\ }\href
  {https://doi.org/10.1021/acs.nanolett.5c03258} {\bibfield  {journal}
  {\bibinfo  {journal} {Nano Lett.}\ }\textbf {\bibinfo {volume} {25}},\
  \bibinfo {pages} {14304} (\bibinfo {year} {2025})}\BibitemShut {NoStop}%
\bibitem [{\citenamefont {Zhang}\ \emph {et~al.}(2014)\citenamefont {Zhang},
  \citenamefont {Liu}, \citenamefont {Luo}, \citenamefont {Freeman},\ and\
  \citenamefont {Zunger}}]{Zhang2014HiddenSpin}%
  \BibitemOpen
  \bibfield  {author} {\bibinfo {author} {\bibfnamefont {X.}~\bibnamefont
  {Zhang}}, \bibinfo {author} {\bibfnamefont {Q.}~\bibnamefont {Liu}}, \bibinfo
  {author} {\bibfnamefont {J.-W.}\ \bibnamefont {Luo}}, \bibinfo {author}
  {\bibfnamefont {A.~J.}\ \bibnamefont {Freeman}},\ and\ \bibinfo {author}
  {\bibfnamefont {A.}~\bibnamefont {Zunger}},\ }\bibfield  {title} {\bibinfo
  {title} {{Hidden spin polarization in inversion-symmetric bulk crystals}},\
  }\href {https://doi.org/10.1038/nphys2933} {\bibfield  {journal} {\bibinfo
  {journal} {Nat. Phys.}\ }\textbf {\bibinfo {volume} {10}},\ \bibinfo {pages}
  {387} (\bibinfo {year} {2014})}\BibitemShut {NoStop}%
\bibitem [{\citenamefont {Guan}\ \emph {et~al.}(2022)\citenamefont {Guan},
  \citenamefont {Xiong}, \citenamefont {Wang},\ and\ \citenamefont
  {Luo}}]{Guan2022ProgressHiddenSpin}%
  \BibitemOpen
  \bibfield  {author} {\bibinfo {author} {\bibfnamefont {S.}~\bibnamefont
  {Guan}}, \bibinfo {author} {\bibfnamefont {J.-X.}\ \bibnamefont {Xiong}},
  \bibinfo {author} {\bibfnamefont {Z.}~\bibnamefont {Wang}},\ and\ \bibinfo
  {author} {\bibfnamefont {J.-W.}\ \bibnamefont {Luo}},\ }\bibfield  {title}
  {\bibinfo {title} {{Progress of hidden spin polarization in
  inversion-symmetric crystals}},\ }\href
  {https://doi.org/10.1007/s11433-021-1821-1} {\bibfield  {journal} {\bibinfo
  {journal} {Sci. China Phys. Mech. Astron.}\ }\textbf {\bibinfo {volume}
  {65}},\ \bibinfo {pages} {237301} (\bibinfo {year} {2022})}\BibitemShut
  {NoStop}%
\bibitem [{\citenamefont {Xiong}\ \emph {et~al.}(2026)\citenamefont {Xiong},
  \citenamefont {Zhang}, \citenamefont {Yuan},\ and\ \citenamefont
  {Zunger}}]{Xiong2026ApparentHidden}%
  \BibitemOpen
  \bibfield  {author} {\bibinfo {author} {\bibfnamefont {J.-X.}\ \bibnamefont
  {Xiong}}, \bibinfo {author} {\bibfnamefont {X.}~\bibnamefont {Zhang}},
  \bibinfo {author} {\bibfnamefont {L.-D.}\ \bibnamefont {Yuan}},\ and\
  \bibinfo {author} {\bibfnamefont {A.}~\bibnamefont {Zunger}},\ }\bibfield
  {title} {\bibinfo {title} {{Matter with apparent and hidden spin physics}},\
  }\href {https://doi.org/10.1016/j.matt.2026.102674} {\bibfield  {journal}
  {\bibinfo  {journal} {Matter}\ }\textbf {\bibinfo {volume} {9}},\ \bibinfo
  {pages} {102674} (\bibinfo {year} {2026})}\BibitemShut {NoStop}%
\bibitem [{\citenamefont {Riley}\ \emph {et~al.}(2014)\citenamefont {Riley},
  \citenamefont {Mazzola}, \citenamefont {Dendzik}, \citenamefont {Michiardi},
  \citenamefont {Takayama}, \citenamefont {Bawden}, \citenamefont
  {Graner{\o}d}, \citenamefont {Leandersson}, \citenamefont {Balasubramanian},
  \citenamefont {Hoesch} \emph {et~al.}}]{Riley2014DirectObservation}%
  \BibitemOpen
  \bibfield  {author} {\bibinfo {author} {\bibfnamefont {J.~M.}\ \bibnamefont
  {Riley}}, \bibinfo {author} {\bibfnamefont {F.}~\bibnamefont {Mazzola}},
  \bibinfo {author} {\bibfnamefont {M.}~\bibnamefont {Dendzik}}, \bibinfo
  {author} {\bibfnamefont {M.}~\bibnamefont {Michiardi}}, \bibinfo {author}
  {\bibfnamefont {T.}~\bibnamefont {Takayama}}, \bibinfo {author}
  {\bibfnamefont {L.}~\bibnamefont {Bawden}}, \bibinfo {author} {\bibfnamefont
  {C.}~\bibnamefont {Graner{\o}d}}, \bibinfo {author} {\bibfnamefont
  {M.}~\bibnamefont {Leandersson}}, \bibinfo {author} {\bibfnamefont
  {T.}~\bibnamefont {Balasubramanian}}, \bibinfo {author} {\bibfnamefont
  {M.}~\bibnamefont {Hoesch}}, \emph {et~al.},\ }\bibfield  {title} {\bibinfo
  {title} {{Direct observation of spin-polarized bulk bands in an
  inversion-symmetric semiconductor}},\ }\href
  {https://doi.org/10.1038/nphys3105} {\bibfield  {journal} {\bibinfo
  {journal} {Nat. Phys.}\ }\textbf {\bibinfo {volume} {10}},\ \bibinfo {pages}
  {835} (\bibinfo {year} {2014})}\BibitemShut {NoStop}%
\bibitem [{\citenamefont {Huang}\ \emph {et~al.}(2020)\citenamefont {Huang},
  \citenamefont {Yartsev}, \citenamefont {Guan}, \citenamefont {Zhu},
  \citenamefont {Zhao}, \citenamefont {Yao}, \citenamefont {He}, \citenamefont
  {Zhang}, \citenamefont {Bai}, \citenamefont {Luo} \emph
  {et~al.}}]{Huang2020HiddenSpin}%
  \BibitemOpen
  \bibfield  {author} {\bibinfo {author} {\bibfnamefont {Y.}~\bibnamefont
  {Huang}}, \bibinfo {author} {\bibfnamefont {A.}~\bibnamefont {Yartsev}},
  \bibinfo {author} {\bibfnamefont {S.}~\bibnamefont {Guan}}, \bibinfo {author}
  {\bibfnamefont {L.}~\bibnamefont {Zhu}}, \bibinfo {author} {\bibfnamefont
  {Q.}~\bibnamefont {Zhao}}, \bibinfo {author} {\bibfnamefont {Z.}~\bibnamefont
  {Yao}}, \bibinfo {author} {\bibfnamefont {C.}~\bibnamefont {He}}, \bibinfo
  {author} {\bibfnamefont {L.}~\bibnamefont {Zhang}}, \bibinfo {author}
  {\bibfnamefont {J.}~\bibnamefont {Bai}}, \bibinfo {author} {\bibfnamefont
  {J.-W.}\ \bibnamefont {Luo}}, \emph {et~al.},\ }\bibfield  {title} {\bibinfo
  {title} {{Hidden spin polarization in the centrosymmetric MoS$_2$ crystal
  revealed via elliptically polarized terahertz emission}},\ }\href
  {https://doi.org/10.1103/PhysRevB.102.085205} {\bibfield  {journal} {\bibinfo
   {journal} {Phys. Rev. B}\ }\textbf {\bibinfo {volume} {102}},\ \bibinfo
  {pages} {085205} (\bibinfo {year} {2020})}\BibitemShut {NoStop}%
\bibitem [{\citenamefont {Zhang}\ \emph {et~al.}(2021)\citenamefont {Zhang},
  \citenamefont {Zhao}, \citenamefont {Hao}, \citenamefont {Kumar},
  \citenamefont {Schwier}, \citenamefont {Zhang}, \citenamefont {Sun},
  \citenamefont {Wang}, \citenamefont {Hao}, \citenamefont {Ma} \emph
  {et~al.}}]{Zhang2021ObservationSpinMomentumLayer}%
  \BibitemOpen
  \bibfield  {author} {\bibinfo {author} {\bibfnamefont {K.}~\bibnamefont
  {Zhang}}, \bibinfo {author} {\bibfnamefont {S.}~\bibnamefont {Zhao}},
  \bibinfo {author} {\bibfnamefont {Z.}~\bibnamefont {Hao}}, \bibinfo {author}
  {\bibfnamefont {S.}~\bibnamefont {Kumar}}, \bibinfo {author} {\bibfnamefont
  {E.~F.}\ \bibnamefont {Schwier}}, \bibinfo {author} {\bibfnamefont
  {Y.}~\bibnamefont {Zhang}}, \bibinfo {author} {\bibfnamefont
  {H.}~\bibnamefont {Sun}}, \bibinfo {author} {\bibfnamefont {Y.}~\bibnamefont
  {Wang}}, \bibinfo {author} {\bibfnamefont {Y.}~\bibnamefont {Hao}}, \bibinfo
  {author} {\bibfnamefont {X.}~\bibnamefont {Ma}}, \emph {et~al.},\ }\bibfield
  {title} {\bibinfo {title} {{Observation of Spin-Momentum-Layer Locking in a
  Centrosymmetric Crystal}},\ }\href
  {https://doi.org/10.1103/PhysRevLett.127.126402} {\bibfield  {journal}
  {\bibinfo  {journal} {Phys. Rev. Lett.}\ }\textbf {\bibinfo {volume} {127}},\
  \bibinfo {pages} {126402} (\bibinfo {year} {2021})}\BibitemShut {NoStop}%
\bibitem [{\citenamefont {Arnoldi}\ \emph {et~al.}(2024)\citenamefont
  {Arnoldi}, \citenamefont {Zachritz}, \citenamefont {Hedwig}, \citenamefont
  {Aeschlimann}, \citenamefont {Monti},\ and\ \citenamefont
  {Stadtm{\"u}ller}}]{Arnoldi2024RevealingHidden}%
  \BibitemOpen
  \bibfield  {author} {\bibinfo {author} {\bibfnamefont {B.}~\bibnamefont
  {Arnoldi}}, \bibinfo {author} {\bibfnamefont {S.~L.}\ \bibnamefont
  {Zachritz}}, \bibinfo {author} {\bibfnamefont {S.}~\bibnamefont {Hedwig}},
  \bibinfo {author} {\bibfnamefont {M.}~\bibnamefont {Aeschlimann}}, \bibinfo
  {author} {\bibfnamefont {O.~L.~A.}\ \bibnamefont {Monti}},\ and\ \bibinfo
  {author} {\bibfnamefont {B.}~\bibnamefont {Stadtm{\"u}ller}},\ }\bibfield
  {title} {\bibinfo {title} {{Revealing hidden spin polarization in
  centrosymmetric van der Waals materials on ultrafast timescales}},\ }\href
  {https://doi.org/10.1038/s41467-024-47821-4} {\bibfield  {journal} {\bibinfo
  {journal} {Nat. Commun.}\ }\textbf {\bibinfo {volume} {15}},\ \bibinfo
  {pages} {3573} (\bibinfo {year} {2024})}\BibitemShut {NoStop}%
\bibitem [{\citenamefont {Okuda}\ \emph {et~al.}(2026)\citenamefont {Okuda},
  \citenamefont {Shishidou}, \citenamefont {Nurmamat}, \citenamefont {Sumida},
  \citenamefont {Schwier}, \citenamefont {Miyamoto},\ and\ \citenamefont
  {Weinert}}]{Okuda2026Hidden}%
  \BibitemOpen
  \bibfield  {author} {\bibinfo {author} {\bibfnamefont {T.}~\bibnamefont
  {Okuda}}, \bibinfo {author} {\bibfnamefont {T.}~\bibnamefont {Shishidou}},
  \bibinfo {author} {\bibfnamefont {M.}~\bibnamefont {Nurmamat}}, \bibinfo
  {author} {\bibfnamefont {K.}~\bibnamefont {Sumida}}, \bibinfo {author}
  {\bibfnamefont {E.}~\bibnamefont {Schwier}}, \bibinfo {author} {\bibfnamefont
  {K.}~\bibnamefont {Miyamoto}},\ and\ \bibinfo {author} {\bibfnamefont
  {M.}~\bibnamefont {Weinert}},\ }\bibfield  {title} {\bibinfo {title}
  {{Unveiling the Hidden Spin-Polarized Bi(110) Surface States by Spin-Resolved
  Photoemission}},\ }\href {https://doi.org/10.1103/3vs9-4y1y} {\bibfield
  {journal} {\bibinfo  {journal} {Phys. Rev. Lett.}\ }\textbf {\bibinfo
  {volume} {137}},\ \bibinfo {pages} {056401} (\bibinfo {year}
  {2026})}\BibitemShut {NoStop}%
\bibitem [{\citenamefont {Chakraborty}\ \emph {et~al.}(2026)\citenamefont
  {Chakraborty}, \citenamefont {Dong}, \citenamefont {Shelton}, \citenamefont
  {Garden}, \citenamefont {Romanetz}, \citenamefont {Hautzinger}, \citenamefont
  {Beard}, \citenamefont {Blum},\ and\ \citenamefont
  {Mitzi}}]{Chakraborty2026HiddenSpinValley}%
  \BibitemOpen
  \bibfield  {author} {\bibinfo {author} {\bibfnamefont {R.}~\bibnamefont
  {Chakraborty}}, \bibinfo {author} {\bibfnamefont {Y.}~\bibnamefont {Dong}},
  \bibinfo {author} {\bibfnamefont {J.~L.}\ \bibnamefont {Shelton}}, \bibinfo
  {author} {\bibfnamefont {K.}~\bibnamefont {Garden}}, \bibinfo {author}
  {\bibfnamefont {L.}~\bibnamefont {Romanetz}}, \bibinfo {author}
  {\bibfnamefont {M.~P.}\ \bibnamefont {Hautzinger}}, \bibinfo {author}
  {\bibfnamefont {M.~C.}\ \bibnamefont {Beard}}, \bibinfo {author}
  {\bibfnamefont {V.}~\bibnamefont {Blum}},\ and\ \bibinfo {author}
  {\bibfnamefont {D.~B.}\ \bibnamefont {Mitzi}},\ }\bibfield  {title} {\bibinfo
  {title} {{Hidden spin-valley locking stabilizes nanosecond spin polarization
  in 2D perovskites}},\ }\bibfield  {journal} {\bibinfo  {journal} {Nat.
  Nanotechnol.}\ }\href {https://doi.org/10.1038/s41565-026-02238-6}
  {10.1038/s41565-026-02238-6} (\bibinfo {year} {2026})\BibitemShut {NoStop}%
\bibitem [{\citenamefont {Galazka}\ \emph {et~al.}(2021)\citenamefont
  {Galazka}, \citenamefont {Irmscher}, \citenamefont {Pietsch}, \citenamefont
  {Ganschow}, \citenamefont {Schulz}, \citenamefont {Klimm}, \citenamefont
  {Hanke}, \citenamefont {Schroeder},\ and\ \citenamefont
  {Bickermann}}]{Galazka2021OxideHallMobility}%
  \BibitemOpen
  \bibfield  {author} {\bibinfo {author} {\bibfnamefont {Z.}~\bibnamefont
  {Galazka}}, \bibinfo {author} {\bibfnamefont {K.}~\bibnamefont {Irmscher}},
  \bibinfo {author} {\bibfnamefont {M.}~\bibnamefont {Pietsch}}, \bibinfo
  {author} {\bibfnamefont {S.}~\bibnamefont {Ganschow}}, \bibinfo {author}
  {\bibfnamefont {D.}~\bibnamefont {Schulz}}, \bibinfo {author} {\bibfnamefont
  {D.}~\bibnamefont {Klimm}}, \bibinfo {author} {\bibfnamefont {I.~M.}\
  \bibnamefont {Hanke}}, \bibinfo {author} {\bibfnamefont {T.}~\bibnamefont
  {Schroeder}},\ and\ \bibinfo {author} {\bibfnamefont {M.}~\bibnamefont
  {Bickermann}},\ }\bibfield  {title} {\bibinfo {title} {Experimental {Hall}
  electron mobility of bulk single crystals of transparent semiconducting
  oxides},\ }\href {https://doi.org/10.1557/s43578-021-00353-9} {\bibfield
  {journal} {\bibinfo  {journal} {J. Mater. Res.}\ }\textbf {\bibinfo {volume}
  {36}},\ \bibinfo {pages} {4746} (\bibinfo {year} {2021})}\BibitemShut
  {NoStop}%
\bibitem [{\citenamefont {Prince}(1953)}]{Prince1953GermaniumMobility}%
  \BibitemOpen
  \bibfield  {author} {\bibinfo {author} {\bibfnamefont {M.~B.}\ \bibnamefont
  {Prince}},\ }\bibfield  {title} {\bibinfo {title} {{Drift Mobilities in
  Semiconductors. I. Germanium}},\ }\href
  {https://doi.org/10.1103/PhysRev.92.681} {\bibfield  {journal} {\bibinfo
  {journal} {Phys. Rev.}\ }\textbf {\bibinfo {volume} {92}},\ \bibinfo {pages}
  {681} (\bibinfo {year} {1953})}\BibitemShut {NoStop}%
\bibitem [{\citenamefont {Segall}\ \emph {et~al.}(1963)\citenamefont {Segall},
  \citenamefont {Lorenz},\ and\ \citenamefont
  {Halsted}}]{Segall1963CdTeElectrical}%
  \BibitemOpen
  \bibfield  {author} {\bibinfo {author} {\bibfnamefont {B.}~\bibnamefont
  {Segall}}, \bibinfo {author} {\bibfnamefont {M.~R.}\ \bibnamefont {Lorenz}},\
  and\ \bibinfo {author} {\bibfnamefont {R.~E.}\ \bibnamefont {Halsted}},\
  }\bibfield  {title} {\bibinfo {title} {{Electrical Properties of {$n$}-Type
  {CdTe}}},\ }\href {https://doi.org/10.1103/PhysRev.129.2471} {\bibfield
  {journal} {\bibinfo  {journal} {Phys. Rev.}\ }\textbf {\bibinfo {volume}
  {129}},\ \bibinfo {pages} {2471} (\bibinfo {year} {1963})}\BibitemShut
  {NoStop}%
\bibitem [{\citenamefont {Jacoboni}\ \emph {et~al.}(1977)\citenamefont
  {Jacoboni}, \citenamefont {Canali}, \citenamefont {Ottaviani},\ and\
  \citenamefont {Alberigi~Quaranta}}]{Jacoboni1977SiliconTransport}%
  \BibitemOpen
  \bibfield  {author} {\bibinfo {author} {\bibfnamefont {C.}~\bibnamefont
  {Jacoboni}}, \bibinfo {author} {\bibfnamefont {C.}~\bibnamefont {Canali}},
  \bibinfo {author} {\bibfnamefont {G.}~\bibnamefont {Ottaviani}},\ and\
  \bibinfo {author} {\bibfnamefont {A.}~\bibnamefont {Alberigi~Quaranta}},\
  }\bibfield  {title} {\bibinfo {title} {A review of some charge transport
  properties of silicon},\ }\href
  {https://doi.org/10.1016/0038-1101(77)90054-5} {\bibfield  {journal}
  {\bibinfo  {journal} {Solid-State Electron.}\ }\textbf {\bibinfo {volume}
  {20}},\ \bibinfo {pages} {77} (\bibinfo {year} {1977})}\BibitemShut {NoStop}%
\bibitem [{\citenamefont {Engh}\ \emph {et~al.}(1981)\citenamefont {Engh},
  \citenamefont {Peterson}, \citenamefont {Thorne},\ and\ \citenamefont
  {Petersen}}]{Engh1981InPCrystal}%
  \BibitemOpen
  \bibfield  {author} {\bibinfo {author} {\bibfnamefont {R.~O.}\ \bibnamefont
  {Engh}}, \bibinfo {author} {\bibfnamefont {S.~R.}\ \bibnamefont {Peterson}},
  \bibinfo {author} {\bibfnamefont {J.~P.}\ \bibnamefont {Thorne}},\ and\
  \bibinfo {author} {\bibfnamefont {P.~E.}\ \bibnamefont {Petersen}},\
  }\bibfield  {title} {\bibinfo {title} {High-purity, single-crystal {InP}
  grown by synthesis solute diffusion},\ }\href
  {https://doi.org/10.1063/1.92330} {\bibfield  {journal} {\bibinfo  {journal}
  {Appl. Phys. Lett.}\ }\textbf {\bibinfo {volume} {38}},\ \bibinfo {pages}
  {243} (\bibinfo {year} {1981})}\BibitemShut {NoStop}%
\bibitem [{\citenamefont {Gu}\ \emph {et~al.}(2016)\citenamefont {Gu},
  \citenamefont {Ren}, \citenamefont {Zhou}, \citenamefont {Tian},
  \citenamefont {Xu}, \citenamefont {Zhang}, \citenamefont {Wang},
  \citenamefont {Wang},\ and\ \citenamefont {Xu}}]{Gu2016GaNElectrical}%
  \BibitemOpen
  \bibfield  {author} {\bibinfo {author} {\bibfnamefont {H.}~\bibnamefont
  {Gu}}, \bibinfo {author} {\bibfnamefont {G.}~\bibnamefont {Ren}}, \bibinfo
  {author} {\bibfnamefont {T.}~\bibnamefont {Zhou}}, \bibinfo {author}
  {\bibfnamefont {F.}~\bibnamefont {Tian}}, \bibinfo {author} {\bibfnamefont
  {Y.}~\bibnamefont {Xu}}, \bibinfo {author} {\bibfnamefont {Y.}~\bibnamefont
  {Zhang}}, \bibinfo {author} {\bibfnamefont {M.}~\bibnamefont {Wang}},
  \bibinfo {author} {\bibfnamefont {J.}~\bibnamefont {Wang}},\ and\ \bibinfo
  {author} {\bibfnamefont {K.}~\bibnamefont {Xu}},\ }\bibfield  {title}
  {\bibinfo {title} {The electrical properties of bulk {GaN} crystals grown by
  {HVPE}},\ }\href {https://doi.org/10.1016/j.jcrysgro.2015.11.027} {\bibfield
  {journal} {\bibinfo  {journal} {J. Cryst. Growth}\ }\textbf {\bibinfo
  {volume} {436}},\ \bibinfo {pages} {76} (\bibinfo {year} {2016})}\BibitemShut
  {NoStop}%
\bibitem [{\citenamefont {Bandurin}\ \emph {et~al.}(2017)\citenamefont
  {Bandurin}, \citenamefont {Tyurnina}, \citenamefont {Yu}, \citenamefont
  {Mishchenko}, \citenamefont {Z{\'o}lyomi}, \citenamefont {Morozov},
  \citenamefont {Kumar}, \citenamefont {Gorbachev}, \citenamefont {Kudrynskyi},
  \citenamefont {Pezzini} \emph {et~al.}}]{Bandurin2017InSeMobility}%
  \BibitemOpen
  \bibfield  {author} {\bibinfo {author} {\bibfnamefont {D.~A.}\ \bibnamefont
  {Bandurin}}, \bibinfo {author} {\bibfnamefont {A.~V.}\ \bibnamefont
  {Tyurnina}}, \bibinfo {author} {\bibfnamefont {G.~L.}\ \bibnamefont {Yu}},
  \bibinfo {author} {\bibfnamefont {A.}~\bibnamefont {Mishchenko}}, \bibinfo
  {author} {\bibfnamefont {V.}~\bibnamefont {Z{\'o}lyomi}}, \bibinfo {author}
  {\bibfnamefont {S.~V.}\ \bibnamefont {Morozov}}, \bibinfo {author}
  {\bibfnamefont {R.~K.}\ \bibnamefont {Kumar}}, \bibinfo {author}
  {\bibfnamefont {R.~V.}\ \bibnamefont {Gorbachev}}, \bibinfo {author}
  {\bibfnamefont {Z.~R.}\ \bibnamefont {Kudrynskyi}}, \bibinfo {author}
  {\bibfnamefont {S.}~\bibnamefont {Pezzini}}, \emph {et~al.},\ }\bibfield
  {title} {\bibinfo {title} {High electron mobility, quantum {Hall} effect and
  anomalous optical response in atomically thin {InSe}},\ }\href
  {https://doi.org/10.1038/nnano.2016.242} {\bibfield  {journal} {\bibinfo
  {journal} {Nat. Nanotechnol.}\ }\textbf {\bibinfo {volume} {12}},\ \bibinfo
  {pages} {223} (\bibinfo {year} {2017})}\BibitemShut {NoStop}%
\bibitem [{\citenamefont {Kimura}\ \emph {et~al.}(2021)\citenamefont {Kimura},
  \citenamefont {Matsumori}, \citenamefont {Oto}, \citenamefont {Kanemitsu},\
  and\ \citenamefont {Yamada}}]{Kimura2021MAPbBr3Mobility}%
  \BibitemOpen
  \bibfield  {author} {\bibinfo {author} {\bibfnamefont {T.}~\bibnamefont
  {Kimura}}, \bibinfo {author} {\bibfnamefont {K.}~\bibnamefont {Matsumori}},
  \bibinfo {author} {\bibfnamefont {K.}~\bibnamefont {Oto}}, \bibinfo {author}
  {\bibfnamefont {Y.}~\bibnamefont {Kanemitsu}},\ and\ \bibinfo {author}
  {\bibfnamefont {Y.}~\bibnamefont {Yamada}},\ }\bibfield  {title} {\bibinfo
  {title} {Observation of high carrier mobility in {CH$_3$NH$_3$PbBr$_3$}
  single crystals by {AC} photo-{Hall} measurements},\ }\href
  {https://doi.org/10.35848/1882-0786/abf02b} {\bibfield  {journal} {\bibinfo
  {journal} {Appl. Phys. Express}\ }\textbf {\bibinfo {volume} {14}},\ \bibinfo
  {pages} {041009} (\bibinfo {year} {2021})}\BibitemShut {NoStop}%
\bibitem [{\citenamefont {Zhao}\ \emph {et~al.}(2024)\citenamefont {Zhao},
  \citenamefont {Pan}, \citenamefont {Chen}, \citenamefont {Cheng},
  \citenamefont {Liu}, \citenamefont {Zhang}, \citenamefont {Zhang},
  \citenamefont {Zhu}, \citenamefont {Song}, \citenamefont {Luo} \emph
  {et~al.}}]{Zhao2024GaSbMobility}%
  \BibitemOpen
  \bibfield  {author} {\bibinfo {author} {\bibfnamefont {Y.}~\bibnamefont
  {Zhao}}, \bibinfo {author} {\bibfnamefont {Y.}~\bibnamefont {Pan}}, \bibinfo
  {author} {\bibfnamefont {L.}~\bibnamefont {Chen}}, \bibinfo {author}
  {\bibfnamefont {M.}~\bibnamefont {Cheng}}, \bibinfo {author} {\bibfnamefont
  {L.}~\bibnamefont {Liu}}, \bibinfo {author} {\bibfnamefont {L.}~\bibnamefont
  {Zhang}}, \bibinfo {author} {\bibfnamefont {R.}~\bibnamefont {Zhang}},
  \bibinfo {author} {\bibfnamefont {X.}~\bibnamefont {Zhu}}, \bibinfo {author}
  {\bibfnamefont {W.}~\bibnamefont {Song}}, \bibinfo {author} {\bibfnamefont
  {X.}~\bibnamefont {Luo}}, \emph {et~al.},\ }\bibfield  {title} {\bibinfo
  {title} {{P}-type electrical transport properties and excellent {Hall}
  mobility of {GaSb} single crystal grown by {Ga} flux},\ }\href
  {https://doi.org/10.1063/5.0232499} {\bibfield  {journal} {\bibinfo
  {journal} {Appl. Phys. Lett.}\ }\textbf {\bibinfo {volume} {125}},\ \bibinfo
  {pages} {262101} (\bibinfo {year} {2024})}\BibitemShut {NoStop}%
\bibitem [{\citenamefont {Wu}\ \emph {et~al.}(2017)\citenamefont {Wu},
  \citenamefont {Yuan}, \citenamefont {Meng}, \citenamefont {Chen},
  \citenamefont {Sun}, \citenamefont {Chen}, \citenamefont {Dang},
  \citenamefont {Tan}, \citenamefont {Liu}, \citenamefont {Yin} \emph
  {et~al.}}]{Wu2017Bi2O2SeMobility}%
  \BibitemOpen
  \bibfield  {author} {\bibinfo {author} {\bibfnamefont {J.}~\bibnamefont
  {Wu}}, \bibinfo {author} {\bibfnamefont {H.}~\bibnamefont {Yuan}}, \bibinfo
  {author} {\bibfnamefont {M.}~\bibnamefont {Meng}}, \bibinfo {author}
  {\bibfnamefont {C.}~\bibnamefont {Chen}}, \bibinfo {author} {\bibfnamefont
  {Y.}~\bibnamefont {Sun}}, \bibinfo {author} {\bibfnamefont {Z.}~\bibnamefont
  {Chen}}, \bibinfo {author} {\bibfnamefont {W.}~\bibnamefont {Dang}}, \bibinfo
  {author} {\bibfnamefont {C.}~\bibnamefont {Tan}}, \bibinfo {author}
  {\bibfnamefont {Y.}~\bibnamefont {Liu}}, \bibinfo {author} {\bibfnamefont
  {J.}~\bibnamefont {Yin}}, \emph {et~al.},\ }\bibfield  {title} {\bibinfo
  {title} {High electron mobility and quantum oscillations in non-encapsulated
  ultrathin semiconducting {Bi$_2$O$_2$Se}},\ }\href
  {https://doi.org/10.1038/nnano.2017.43} {\bibfield  {journal} {\bibinfo
  {journal} {Nat. Nanotechnol.}\ }\textbf {\bibinfo {volume} {12}},\ \bibinfo
  {pages} {530} (\bibinfo {year} {2017})}\BibitemShut {NoStop}%
\bibitem [{\citenamefont {Radisavljevic}\ \emph {et~al.}(2011)\citenamefont
  {Radisavljevic}, \citenamefont {Radenovic}, \citenamefont {Brivio},
  \citenamefont {Giacometti},\ and\ \citenamefont
  {Kis}}]{Radisavljevic2011MoS2Transistors}%
  \BibitemOpen
  \bibfield  {author} {\bibinfo {author} {\bibfnamefont {B.}~\bibnamefont
  {Radisavljevic}}, \bibinfo {author} {\bibfnamefont {A.}~\bibnamefont
  {Radenovic}}, \bibinfo {author} {\bibfnamefont {J.}~\bibnamefont {Brivio}},
  \bibinfo {author} {\bibfnamefont {V.}~\bibnamefont {Giacometti}},\ and\
  \bibinfo {author} {\bibfnamefont {A.}~\bibnamefont {Kis}},\ }\bibfield
  {title} {\bibinfo {title} {Single-layer {MoS$_2$} transistors},\ }\href
  {https://doi.org/10.1038/nnano.2010.279} {\bibfield  {journal} {\bibinfo
  {journal} {Nat. Nanotechnol.}\ }\textbf {\bibinfo {volume} {6}},\ \bibinfo
  {pages} {147} (\bibinfo {year} {2011})}\BibitemShut {NoStop}%
\bibitem [{\citenamefont {Pack}\ \emph {et~al.}(2024)\citenamefont {Pack},
  \citenamefont {Guo}, \citenamefont {Liu}, \citenamefont {Jessen},
  \citenamefont {Holtzman}, \citenamefont {Liu}, \citenamefont {Cothrine},
  \citenamefont {Watanabe}, \citenamefont {Taniguchi}, \citenamefont {Mandrus}
  \emph {et~al.}}]{Pack2024WSe2ChargeTransferContacts}%
  \BibitemOpen
  \bibfield  {author} {\bibinfo {author} {\bibfnamefont {J.}~\bibnamefont
  {Pack}}, \bibinfo {author} {\bibfnamefont {Y.}~\bibnamefont {Guo}}, \bibinfo
  {author} {\bibfnamefont {Z.}~\bibnamefont {Liu}}, \bibinfo {author}
  {\bibfnamefont {B.~S.}\ \bibnamefont {Jessen}}, \bibinfo {author}
  {\bibfnamefont {L.}~\bibnamefont {Holtzman}}, \bibinfo {author}
  {\bibfnamefont {S.}~\bibnamefont {Liu}}, \bibinfo {author} {\bibfnamefont
  {M.}~\bibnamefont {Cothrine}}, \bibinfo {author} {\bibfnamefont
  {K.}~\bibnamefont {Watanabe}}, \bibinfo {author} {\bibfnamefont
  {T.}~\bibnamefont {Taniguchi}}, \bibinfo {author} {\bibfnamefont {D.~G.}\
  \bibnamefont {Mandrus}}, \emph {et~al.},\ }\bibfield  {title} {\bibinfo
  {title} {Charge-transfer contacts for the measurement of correlated states in
  high-mobility {WSe$_2$}},\ }\href
  {https://doi.org/10.1038/s41565-024-01702-5} {\bibfield  {journal} {\bibinfo
  {journal} {Nat. Nanotechnol.}\ }\textbf {\bibinfo {volume} {19}},\ \bibinfo
  {pages} {948} (\bibinfo {year} {2024})}\BibitemShut {NoStop}%
\bibitem [{\citenamefont {Wang}\ \emph {et~al.}(2009)\citenamefont {Wang},
  \citenamefont {Li}, \citenamefont {Li}, \citenamefont {Xu}, \citenamefont
  {Cui}, \citenamefont {Oganov},\ and\ \citenamefont
  {Ma}}]{Wang2009Ultraincompressible}%
  \BibitemOpen
  \bibfield  {author} {\bibinfo {author} {\bibfnamefont {H.}~\bibnamefont
  {Wang}}, \bibinfo {author} {\bibfnamefont {Q.}~\bibnamefont {Li}}, \bibinfo
  {author} {\bibfnamefont {Y.}~\bibnamefont {Li}}, \bibinfo {author}
  {\bibfnamefont {Y.}~\bibnamefont {Xu}}, \bibinfo {author} {\bibfnamefont
  {T.}~\bibnamefont {Cui}}, \bibinfo {author} {\bibfnamefont {A.~R.}\
  \bibnamefont {Oganov}},\ and\ \bibinfo {author} {\bibfnamefont
  {Y.}~\bibnamefont {Ma}},\ }\bibfield  {title} {\bibinfo {title}
  {{Ultra-incompressible phases of tungsten dinitride predicted from first
  principles}},\ }\href {https://doi.org/10.1103/PhysRevB.79.132109} {\bibfield
   {journal} {\bibinfo  {journal} {Phys. Rev. B}\ }\textbf {\bibinfo {volume}
  {79}},\ \bibinfo {pages} {132109} (\bibinfo {year} {2009})}\BibitemShut
  {NoStop}%
\bibitem [{\citenamefont {Perdew}\ \emph {et~al.}(2008)\citenamefont {Perdew},
  \citenamefont {Ruzsinszky}, \citenamefont {Csonka}, \citenamefont {Vydrov},
  \citenamefont {Scuseria}, \citenamefont {Constantin}, \citenamefont {Zhou},\
  and\ \citenamefont {Burke}}]{Perdew2008Restoring}%
  \BibitemOpen
  \bibfield  {author} {\bibinfo {author} {\bibfnamefont {J.~P.}\ \bibnamefont
  {Perdew}}, \bibinfo {author} {\bibfnamefont {A.}~\bibnamefont {Ruzsinszky}},
  \bibinfo {author} {\bibfnamefont {G.~I.}\ \bibnamefont {Csonka}}, \bibinfo
  {author} {\bibfnamefont {O.~A.}\ \bibnamefont {Vydrov}}, \bibinfo {author}
  {\bibfnamefont {G.~E.}\ \bibnamefont {Scuseria}}, \bibinfo {author}
  {\bibfnamefont {L.~A.}\ \bibnamefont {Constantin}}, \bibinfo {author}
  {\bibfnamefont {X.}~\bibnamefont {Zhou}},\ and\ \bibinfo {author}
  {\bibfnamefont {K.}~\bibnamefont {Burke}},\ }\bibfield  {title} {\bibinfo
  {title} {{Restoring the Density-Gradient Expansion for Exchange in Solids and
  Surfaces}},\ }\href {https://doi.org/10.1103/PhysRevLett.100.136406}
  {\bibfield  {journal} {\bibinfo  {journal} {Phys. Rev. Lett.}\ }\textbf
  {\bibinfo {volume} {100}},\ \bibinfo {pages} {136406} (\bibinfo {year}
  {2008})}\BibitemShut {NoStop}%
\bibitem [{\citenamefont {Giannozzi}\ \emph {et~al.}(2017)\citenamefont
  {Giannozzi}, \citenamefont {Andreussi}, \citenamefont {Brumme}, \citenamefont
  {Bunau}, \citenamefont {Buongiorno~Nardelli}, \citenamefont {Calandra},
  \citenamefont {Car}, \citenamefont {Cavazzoni}, \citenamefont {Ceresoli},
  \citenamefont {Cococcioni} \emph {et~al.}}]{Giannozzi2017Advanced}%
  \BibitemOpen
  \bibfield  {author} {\bibinfo {author} {\bibfnamefont {P.}~\bibnamefont
  {Giannozzi}}, \bibinfo {author} {\bibfnamefont {O.}~\bibnamefont
  {Andreussi}}, \bibinfo {author} {\bibfnamefont {T.}~\bibnamefont {Brumme}},
  \bibinfo {author} {\bibfnamefont {O.}~\bibnamefont {Bunau}}, \bibinfo
  {author} {\bibfnamefont {M.}~\bibnamefont {Buongiorno~Nardelli}}, \bibinfo
  {author} {\bibfnamefont {M.}~\bibnamefont {Calandra}}, \bibinfo {author}
  {\bibfnamefont {R.}~\bibnamefont {Car}}, \bibinfo {author} {\bibfnamefont
  {C.}~\bibnamefont {Cavazzoni}}, \bibinfo {author} {\bibfnamefont
  {D.}~\bibnamefont {Ceresoli}}, \bibinfo {author} {\bibfnamefont
  {M.}~\bibnamefont {Cococcioni}}, \emph {et~al.},\ }\bibfield  {title}
  {\bibinfo {title} {{Advanced Capabilities for Materials Modelling with
  {QUANTUM ESPRESSO}}},\ }\href {https://doi.org/10.1088/1361-648X/aa8f79}
  {\bibfield  {journal} {\bibinfo  {journal} {J. Phys.: Condens. Matter}\
  }\textbf {\bibinfo {volume} {29}},\ \bibinfo {pages} {465901} (\bibinfo
  {year} {2017})}\BibitemShut {NoStop}%
\bibitem [{\citenamefont {Krukau}\ \emph {et~al.}(2006)\citenamefont {Krukau},
  \citenamefont {Vydrov}, \citenamefont {Izmaylov},\ and\ \citenamefont
  {Scuseria}}]{Krukau2006Influence}%
  \BibitemOpen
  \bibfield  {author} {\bibinfo {author} {\bibfnamefont {A.~V.}\ \bibnamefont
  {Krukau}}, \bibinfo {author} {\bibfnamefont {O.~A.}\ \bibnamefont {Vydrov}},
  \bibinfo {author} {\bibfnamefont {A.~F.}\ \bibnamefont {Izmaylov}},\ and\
  \bibinfo {author} {\bibfnamefont {G.~E.}\ \bibnamefont {Scuseria}},\
  }\bibfield  {title} {\bibinfo {title} {{Influence of the Exchange Screening
  Parameter on the Performance of Screened Hybrid Functionals}},\ }\href
  {https://doi.org/10.1063/1.2404663} {\bibfield  {journal} {\bibinfo
  {journal} {J. Chem. Phys.}\ }\textbf {\bibinfo {volume} {125}},\ \bibinfo
  {pages} {224106} (\bibinfo {year} {2006})}\BibitemShut {NoStop}%
\bibitem [{\citenamefont {Kresse}\ and\ \citenamefont
  {Furthm{\"u}ller}(1996)}]{Kresse1996Efficient}%
  \BibitemOpen
  \bibfield  {author} {\bibinfo {author} {\bibfnamefont {G.}~\bibnamefont
  {Kresse}}\ and\ \bibinfo {author} {\bibfnamefont {J.}~\bibnamefont
  {Furthm{\"u}ller}},\ }\bibfield  {title} {\bibinfo {title} {{Efficient
  Iterative Schemes for Ab Initio Total-Energy Calculations Using a Plane-Wave
  Basis Set}},\ }\href {https://doi.org/10.1103/PhysRevB.54.11169} {\bibfield
  {journal} {\bibinfo  {journal} {Phys. Rev. B}\ }\textbf {\bibinfo {volume}
  {54}},\ \bibinfo {pages} {11169} (\bibinfo {year} {1996})}\BibitemShut
  {NoStop}%
\bibitem [{\citenamefont {Lee}\ \emph {et~al.}(2018)\citenamefont {Lee},
  \citenamefont {Zhou}, \citenamefont {Agapito},\ and\ \citenamefont
  {Bernardi}}]{Lee2018naphthalene}%
  \BibitemOpen
  \bibfield  {author} {\bibinfo {author} {\bibfnamefont {N.-E.}\ \bibnamefont
  {Lee}}, \bibinfo {author} {\bibfnamefont {J.-J.}\ \bibnamefont {Zhou}},
  \bibinfo {author} {\bibfnamefont {L.~A.}\ \bibnamefont {Agapito}},\ and\
  \bibinfo {author} {\bibfnamefont {M.}~\bibnamefont {Bernardi}},\ }\bibfield
  {title} {\bibinfo {title} {Charge transport in organic molecular
  semiconductors from first principles: {The} bandlike hole mobility in a
  naphthalene crystal},\ }\href {https://doi.org/10.1103/PhysRevB.97.115203}
  {\bibfield  {journal} {\bibinfo  {journal} {Phys. Rev. B}\ }\textbf {\bibinfo
  {volume} {97}},\ \bibinfo {pages} {115203} (\bibinfo {year}
  {2018})}\BibitemShut {NoStop}%
\bibitem [{\citenamefont {Abramovitch}\ \emph {et~al.}(2023)\citenamefont
  {Abramovitch}, \citenamefont {Zhou}, \citenamefont {Mravlje}, \citenamefont
  {Georges},\ and\ \citenamefont {Bernardi}}]{Abramovitch2023}%
  \BibitemOpen
  \bibfield  {author} {\bibinfo {author} {\bibfnamefont {D.~J.}\ \bibnamefont
  {Abramovitch}}, \bibinfo {author} {\bibfnamefont {J.-J.}\ \bibnamefont
  {Zhou}}, \bibinfo {author} {\bibfnamefont {J.}~\bibnamefont {Mravlje}},
  \bibinfo {author} {\bibfnamefont {A.}~\bibnamefont {Georges}},\ and\ \bibinfo
  {author} {\bibfnamefont {M.}~\bibnamefont {Bernardi}},\ }\bibfield  {title}
  {\bibinfo {title} {Combining electron-phonon and dynamical mean-field theory
  calculations of correlated materials: {Transport} in the correlated metal
  {Sr}$_{2}${RuO}$_{4}$},\ }\href
  {https://doi.org/10.1103/PhysRevMaterials.7.093801} {\bibfield  {journal}
  {\bibinfo  {journal} {Phys. Rev. Mater.}\ }\textbf {\bibinfo {volume} {7}},\
  \bibinfo {pages} {093801} (\bibinfo {year} {2023})}\BibitemShut {NoStop}%
\bibitem [{\citenamefont {Hu}\ \emph {et~al.}(2026)\citenamefont {Hu},
  \citenamefont {Gong}, \citenamefont {Qiu}, \citenamefont {Yang},
  \citenamefont {Zhou},\ and\ \citenamefont {Yao}}]{Hu2026Phonon}%
  \BibitemOpen
  \bibfield  {author} {\bibinfo {author} {\bibfnamefont {W.}~\bibnamefont
  {Hu}}, \bibinfo {author} {\bibfnamefont {J.}~\bibnamefont {Gong}}, \bibinfo
  {author} {\bibfnamefont {Y.}~\bibnamefont {Qiu}}, \bibinfo {author}
  {\bibfnamefont {L.}~\bibnamefont {Yang}}, \bibinfo {author} {\bibfnamefont
  {J.-J.}\ \bibnamefont {Zhou}},\ and\ \bibinfo {author} {\bibfnamefont
  {Y.}~\bibnamefont {Yao}},\ }\bibfield  {title} {\bibinfo {title} {Phonons
  drive the topological phase transition in quasi-one-dimensional
  {Bi}$_{4}${I}$_{4}$},\ }\href {https://doi.org/10.1103/dpwc-vt44} {\bibfield
  {journal} {\bibinfo  {journal} {Phys. Rev. B}\ }\textbf {\bibinfo {volume}
  {113}},\ \bibinfo {pages} {L201111} (\bibinfo {year} {2026})}\BibitemShut
  {NoStop}%
\bibitem [{\citenamefont {Zhou}\ \emph {et~al.}(2021)\citenamefont {Zhou},
  \citenamefont {Park}, \citenamefont {Lu}, \citenamefont {Maliyov},
  \citenamefont {Tong},\ and\ \citenamefont {Bernardi}}]{Zhou2021Perturbo}%
  \BibitemOpen
  \bibfield  {author} {\bibinfo {author} {\bibfnamefont {J.-J.}\ \bibnamefont
  {Zhou}}, \bibinfo {author} {\bibfnamefont {J.}~\bibnamefont {Park}}, \bibinfo
  {author} {\bibfnamefont {I.-T.}\ \bibnamefont {Lu}}, \bibinfo {author}
  {\bibfnamefont {I.}~\bibnamefont {Maliyov}}, \bibinfo {author} {\bibfnamefont
  {X.}~\bibnamefont {Tong}},\ and\ \bibinfo {author} {\bibfnamefont
  {M.}~\bibnamefont {Bernardi}},\ }\bibfield  {title} {\bibinfo {title}
  {{Perturbo: A software package for ab initio electron--phonon interactions,
  charge transport and ultrafast dynamics}},\ }\href
  {https://doi.org/10.1016/j.cpc.2021.107970} {\bibfield  {journal} {\bibinfo
  {journal} {Comput. Phys. Commun.}\ }\textbf {\bibinfo {volume} {264}},\
  \bibinfo {pages} {107970} (\bibinfo {year} {2021})}\BibitemShut {NoStop}%
\bibitem [{\citenamefont {Jhalani}\ \emph {et~al.}(2020)\citenamefont
  {Jhalani}, \citenamefont {Zhou}, \citenamefont {Park}, \citenamefont
  {Dreyer},\ and\ \citenamefont {Bernardi}}]{Jhalani2020PiezoelectricElectron}%
  \BibitemOpen
  \bibfield  {author} {\bibinfo {author} {\bibfnamefont {V.~A.}\ \bibnamefont
  {Jhalani}}, \bibinfo {author} {\bibfnamefont {J.-J.}\ \bibnamefont {Zhou}},
  \bibinfo {author} {\bibfnamefont {J.}~\bibnamefont {Park}}, \bibinfo {author}
  {\bibfnamefont {C.~E.}\ \bibnamefont {Dreyer}},\ and\ \bibinfo {author}
  {\bibfnamefont {M.}~\bibnamefont {Bernardi}},\ }\bibfield  {title} {\bibinfo
  {title} {{Piezoelectric Electron-Phonon Interaction from Ab Initio Dynamical
  Quadrupoles: Impact on Charge Transport in Wurtzite GaN}},\ }\href
  {https://doi.org/10.1103/PhysRevLett.125.136602} {\bibfield  {journal}
  {\bibinfo  {journal} {Phys. Rev. Lett.}\ }\textbf {\bibinfo {volume} {125}},\
  \bibinfo {pages} {136602} (\bibinfo {year} {2020})}\BibitemShut {NoStop}%
\bibitem [{\citenamefont {Park}\ \emph {et~al.}(2020)\citenamefont {Park},
  \citenamefont {Zhou}, \citenamefont {Jhalani}, \citenamefont {Dreyer},\ and\
  \citenamefont {Bernardi}}]{Park2020LongRangeQuadrupole}%
  \BibitemOpen
  \bibfield  {author} {\bibinfo {author} {\bibfnamefont {J.}~\bibnamefont
  {Park}}, \bibinfo {author} {\bibfnamefont {J.-J.}\ \bibnamefont {Zhou}},
  \bibinfo {author} {\bibfnamefont {V.~A.}\ \bibnamefont {Jhalani}}, \bibinfo
  {author} {\bibfnamefont {C.~E.}\ \bibnamefont {Dreyer}},\ and\ \bibinfo
  {author} {\bibfnamefont {M.}~\bibnamefont {Bernardi}},\ }\bibfield  {title}
  {\bibinfo {title} {{Long-range quadrupole electron-phonon interaction from
  first principles}},\ }\href {https://doi.org/10.1103/PhysRevB.102.125203}
  {\bibfield  {journal} {\bibinfo  {journal} {Phys. Rev. B}\ }\textbf {\bibinfo
  {volume} {102}},\ \bibinfo {pages} {125203} (\bibinfo {year}
  {2020})}\BibitemShut {NoStop}%
\bibitem [{\citenamefont {Brunin}\ \emph
  {et~al.}(2020{\natexlab{a}})\citenamefont {Brunin}, \citenamefont {Miranda},
  \citenamefont {Giantomassi}, \citenamefont {Royo}, \citenamefont {Stengel},
  \citenamefont {Verstraete}, \citenamefont {Gonze}, \citenamefont
  {Rignanese},\ and\ \citenamefont {Hautier}}]{Brunin2020PRL}%
  \BibitemOpen
  \bibfield  {author} {\bibinfo {author} {\bibfnamefont {G.}~\bibnamefont
  {Brunin}}, \bibinfo {author} {\bibfnamefont {H.~P.~C.}\ \bibnamefont
  {Miranda}}, \bibinfo {author} {\bibfnamefont {M.}~\bibnamefont
  {Giantomassi}}, \bibinfo {author} {\bibfnamefont {M.}~\bibnamefont {Royo}},
  \bibinfo {author} {\bibfnamefont {M.}~\bibnamefont {Stengel}}, \bibinfo
  {author} {\bibfnamefont {M.~J.}\ \bibnamefont {Verstraete}}, \bibinfo
  {author} {\bibfnamefont {X.}~\bibnamefont {Gonze}}, \bibinfo {author}
  {\bibfnamefont {G.-M.}\ \bibnamefont {Rignanese}},\ and\ \bibinfo {author}
  {\bibfnamefont {G.}~\bibnamefont {Hautier}},\ }\bibfield  {title} {\bibinfo
  {title} {{Electron-Phonon beyond Fr\"ohlich: Dynamical Quadrupoles in Polar
  and Covalent Solids}},\ }\href
  {https://doi.org/10.1103/PhysRevLett.125.136601} {\bibfield  {journal}
  {\bibinfo  {journal} {Phys. Rev. Lett.}\ }\textbf {\bibinfo {volume} {125}},\
  \bibinfo {pages} {136601} (\bibinfo {year} {2020}{\natexlab{a}})}\BibitemShut
  {NoStop}%
\bibitem [{\citenamefont {Brunin}\ \emph
  {et~al.}(2020{\natexlab{b}})\citenamefont {Brunin}, \citenamefont {Miranda},
  \citenamefont {Giantomassi}, \citenamefont {Royo}, \citenamefont {Stengel},
  \citenamefont {Verstraete}, \citenamefont {Gonze}, \citenamefont
  {Rignanese},\ and\ \citenamefont {Hautier}}]{Brunin2020PRB}%
  \BibitemOpen
  \bibfield  {author} {\bibinfo {author} {\bibfnamefont {G.}~\bibnamefont
  {Brunin}}, \bibinfo {author} {\bibfnamefont {H.~P.~C.}\ \bibnamefont
  {Miranda}}, \bibinfo {author} {\bibfnamefont {M.}~\bibnamefont
  {Giantomassi}}, \bibinfo {author} {\bibfnamefont {M.}~\bibnamefont {Royo}},
  \bibinfo {author} {\bibfnamefont {M.}~\bibnamefont {Stengel}}, \bibinfo
  {author} {\bibfnamefont {M.~J.}\ \bibnamefont {Verstraete}}, \bibinfo
  {author} {\bibfnamefont {X.}~\bibnamefont {Gonze}}, \bibinfo {author}
  {\bibfnamefont {G.-M.}\ \bibnamefont {Rignanese}},\ and\ \bibinfo {author}
  {\bibfnamefont {G.}~\bibnamefont {Hautier}},\ }\bibfield  {title} {\bibinfo
  {title} {Phonon-limited electron mobility in {Si}, {GaAs}, and {GaP} with
  exact treatment of dynamical quadrupoles},\ }\href
  {https://doi.org/10.1103/PhysRevB.102.094308} {\bibfield  {journal} {\bibinfo
   {journal} {Phys. Rev. B}\ }\textbf {\bibinfo {volume} {102}},\ \bibinfo
  {pages} {094308} (\bibinfo {year} {2020}{\natexlab{b}})}\BibitemShut
  {NoStop}%
\bibitem [{\citenamefont {Sjakste}\ \emph {et~al.}(2015)\citenamefont
  {Sjakste}, \citenamefont {Vast}, \citenamefont {Calandra},\ and\
  \citenamefont {Mauri}}]{Sjakste2015}%
  \BibitemOpen
  \bibfield  {author} {\bibinfo {author} {\bibfnamefont {J.}~\bibnamefont
  {Sjakste}}, \bibinfo {author} {\bibfnamefont {N.}~\bibnamefont {Vast}},
  \bibinfo {author} {\bibfnamefont {M.}~\bibnamefont {Calandra}},\ and\
  \bibinfo {author} {\bibfnamefont {F.}~\bibnamefont {Mauri}},\ }\bibfield
  {title} {\bibinfo {title} {Wannier interpolation of the electron-phonon
  matrix elements in polar semiconductors: {Polar-optical} coupling in
  {GaAs}},\ }\href {https://doi.org/10.1103/PhysRevB.92.054307} {\bibfield
  {journal} {\bibinfo  {journal} {Phys. Rev. B}\ }\textbf {\bibinfo {volume}
  {92}},\ \bibinfo {pages} {054307} (\bibinfo {year} {2015})}\BibitemShut
  {NoStop}%
\bibitem [{\citenamefont {Verdi}\ and\ \citenamefont
  {Giustino}(2015)}]{Verdi2015}%
  \BibitemOpen
  \bibfield  {author} {\bibinfo {author} {\bibfnamefont {C.}~\bibnamefont
  {Verdi}}\ and\ \bibinfo {author} {\bibfnamefont {F.}~\bibnamefont
  {Giustino}},\ }\bibfield  {title} {\bibinfo {title} {{Fr\"ohlich
  Electron-Phonon Vertex from First Principles}},\ }\href
  {https://doi.org/10.1103/PhysRevLett.115.176401} {\bibfield  {journal}
  {\bibinfo  {journal} {Phys. Rev. Lett.}\ }\textbf {\bibinfo {volume} {115}},\
  \bibinfo {pages} {176401} (\bibinfo {year} {2015})}\BibitemShut {NoStop}%
\bibitem [{\citenamefont {Baroni}\ \emph {et~al.}(2001)\citenamefont {Baroni},
  \citenamefont {de~Gironcoli}, \citenamefont {Dal~Corso},\ and\ \citenamefont
  {Giannozzi}}]{Baroni2001RMP}%
  \BibitemOpen
  \bibfield  {author} {\bibinfo {author} {\bibfnamefont {S.}~\bibnamefont
  {Baroni}}, \bibinfo {author} {\bibfnamefont {S.}~\bibnamefont
  {de~Gironcoli}}, \bibinfo {author} {\bibfnamefont {A.}~\bibnamefont
  {Dal~Corso}},\ and\ \bibinfo {author} {\bibfnamefont {P.}~\bibnamefont
  {Giannozzi}},\ }\bibfield  {title} {\bibinfo {title} {Phonons and related
  crystal properties from density-functional perturbation theory},\ }\href
  {https://doi.org/10.1103/RevModPhys.73.515} {\bibfield  {journal} {\bibinfo
  {journal} {Rev. Mod. Phys.}\ }\textbf {\bibinfo {volume} {73}},\ \bibinfo
  {pages} {515} (\bibinfo {year} {2001})}\BibitemShut {NoStop}%
\bibitem [{Sup()}]{SupplementalMaterial}%
  \BibitemOpen
  \href@noop {} {}\bibinfo {note} {See {Supplemental Material} at link for
  details on the computational methods, phonon dispersions and band-edge
  properties, validation of electron--phonon interpolation, the formula for
  thermally averaged scattering rates, sector-resolved spin polarization and
  intervalley selection rules, and long-range dipole and quadrupole responses,
  which includes Refs.~\cite{Hamann2013Optimized, vanSetten2018pseudodojo,
  Mostofi2008Wannier90, Gonze2020, Aroyo2006Bilbao, Elcoro2017Double,
  KingSmith1993Polarization}}\BibitemShut {NoStop}%
\bibitem [{\citenamefont {Liu}\ \emph {et~al.}(2013)\citenamefont {Liu},
  \citenamefont {Shan}, \citenamefont {Yao}, \citenamefont {Yao},\ and\
  \citenamefont {Xiao}}]{Liu2013ThreeBand}%
  \BibitemOpen
  \bibfield  {author} {\bibinfo {author} {\bibfnamefont {G.-B.}\ \bibnamefont
  {Liu}}, \bibinfo {author} {\bibfnamefont {W.-Y.}\ \bibnamefont {Shan}},
  \bibinfo {author} {\bibfnamefont {Y.}~\bibnamefont {Yao}}, \bibinfo {author}
  {\bibfnamefont {W.}~\bibnamefont {Yao}},\ and\ \bibinfo {author}
  {\bibfnamefont {D.}~\bibnamefont {Xiao}},\ }\bibfield  {title} {\bibinfo
  {title} {Three-band tight-binding model for monolayers of group-vib
  transition metal dichalcogenides},\ }\href
  {https://doi.org/10.1103/PhysRevB.88.085433} {\bibfield  {journal} {\bibinfo
  {journal} {Phys. Rev. B}\ }\textbf {\bibinfo {volume} {88}},\ \bibinfo
  {pages} {085433} (\bibinfo {year} {2013})}\BibitemShut {NoStop}%
\bibitem [{\citenamefont {Guan}\ \emph {et~al.}(2023)\citenamefont {Guan},
  \citenamefont {Luo}, \citenamefont {Li},\ and\ \citenamefont
  {Zunger}}]{Guan2023HiddenZeeman}%
  \BibitemOpen
  \bibfield  {author} {\bibinfo {author} {\bibfnamefont {S.}~\bibnamefont
  {Guan}}, \bibinfo {author} {\bibfnamefont {J.-W.}\ \bibnamefont {Luo}},
  \bibinfo {author} {\bibfnamefont {S.-S.}\ \bibnamefont {Li}},\ and\ \bibinfo
  {author} {\bibfnamefont {A.}~\bibnamefont {Zunger}},\ }\bibfield  {title}
  {\bibinfo {title} {{Hidden Zeeman-type spin polarization in bulk crystals}},\
  }\href {https://doi.org/10.1103/PhysRevB.107.L081201} {\bibfield  {journal}
  {\bibinfo  {journal} {Phys. Rev. B}\ }\textbf {\bibinfo {volume} {107}},\
  \bibinfo {pages} {L081201} (\bibinfo {year} {2023})}\BibitemShut {NoStop}%
\bibitem [{\citenamefont {Wu}\ \emph {et~al.}(2018)\citenamefont {Wu},
  \citenamefont {Liu}, \citenamefont {Li}, \citenamefont {Zhong}, \citenamefont
  {Yu}, \citenamefont {Sheng}, \citenamefont {Zhao},\ and\ \citenamefont
  {Yang}}]{Wu2018nodal}%
  \BibitemOpen
  \bibfield  {author} {\bibinfo {author} {\bibfnamefont {W.}~\bibnamefont
  {Wu}}, \bibinfo {author} {\bibfnamefont {Y.}~\bibnamefont {Liu}}, \bibinfo
  {author} {\bibfnamefont {S.}~\bibnamefont {Li}}, \bibinfo {author}
  {\bibfnamefont {C.}~\bibnamefont {Zhong}}, \bibinfo {author} {\bibfnamefont
  {Z.-M.}\ \bibnamefont {Yu}}, \bibinfo {author} {\bibfnamefont {X.-L.}\
  \bibnamefont {Sheng}}, \bibinfo {author} {\bibfnamefont {Y.~X.}\ \bibnamefont
  {Zhao}},\ and\ \bibinfo {author} {\bibfnamefont {S.~A.}\ \bibnamefont
  {Yang}},\ }\bibfield  {title} {\bibinfo {title} {Nodal surface semimetals:
  {Theory} and material realization},\ }\href
  {https://doi.org/10.1103/PhysRevB.97.115125} {\bibfield  {journal} {\bibinfo
  {journal} {Phys. Rev. B}\ }\textbf {\bibinfo {volume} {97}},\ \bibinfo
  {pages} {115125} (\bibinfo {year} {2018})}\BibitemShut {NoStop}%
\bibitem [{\citenamefont {Zhang}\ \emph {et~al.}(2025)\citenamefont {Zhang},
  \citenamefont {Cheng}, \citenamefont {Yin}, \citenamefont {Liu},
  \citenamefont {Deng}, \citenamefont {Qiao}, \citenamefont {Shi},
  \citenamefont {Zhang}, \citenamefont {Lin}, \citenamefont {Liu} \emph
  {et~al.}}]{Zhang2025CrystalSymmetryPaired}%
  \BibitemOpen
  \bibfield  {author} {\bibinfo {author} {\bibfnamefont {F.}~\bibnamefont
  {Zhang}}, \bibinfo {author} {\bibfnamefont {X.}~\bibnamefont {Cheng}},
  \bibinfo {author} {\bibfnamefont {Z.}~\bibnamefont {Yin}}, \bibinfo {author}
  {\bibfnamefont {C.}~\bibnamefont {Liu}}, \bibinfo {author} {\bibfnamefont
  {L.}~\bibnamefont {Deng}}, \bibinfo {author} {\bibfnamefont {Y.}~\bibnamefont
  {Qiao}}, \bibinfo {author} {\bibfnamefont {Z.}~\bibnamefont {Shi}}, \bibinfo
  {author} {\bibfnamefont {S.}~\bibnamefont {Zhang}}, \bibinfo {author}
  {\bibfnamefont {J.}~\bibnamefont {Lin}}, \bibinfo {author} {\bibfnamefont
  {Z.}~\bibnamefont {Liu}}, \emph {et~al.},\ }\bibfield  {title} {\bibinfo
  {title} {{Crystal-symmetry-paired spin--valley locking in a layered
  room-temperature metallic altermagnet candidate}},\ }\href
  {https://doi.org/10.1038/s41567-025-02864-2} {\bibfield  {journal} {\bibinfo
  {journal} {Nat. Phys.}\ }\textbf {\bibinfo {volume} {21}},\ \bibinfo {pages}
  {760} (\bibinfo {year} {2025})}\BibitemShut {NoStop}%
\bibitem [{dat()}]{data}%
  \BibitemOpen
  \href {https://doi.org/10.6084/m9.figshare.33154703} {}\bibinfo
  {howpublished}
  {\url{https://doi.org/10.6084/m9.figshare.33154703}}\BibitemShut {NoStop}%
\bibitem [{\citenamefont {Hamann}(2013)}]{Hamann2013Optimized}%
  \BibitemOpen
  \bibfield  {author} {\bibinfo {author} {\bibfnamefont {D.~R.}\ \bibnamefont
  {Hamann}},\ }\bibfield  {title} {\bibinfo {title} {{Optimized Norm-Conserving
  Vanderbilt Pseudopotentials}},\ }\href
  {https://doi.org/10.1103/PhysRevB.88.085117} {\bibfield  {journal} {\bibinfo
  {journal} {Phys. Rev. B}\ }\textbf {\bibinfo {volume} {88}},\ \bibinfo
  {pages} {085117} (\bibinfo {year} {2013})}\BibitemShut {NoStop}%
\bibitem [{\citenamefont {van Setten}\ \emph {et~al.}(2018)\citenamefont {van
  Setten}, \citenamefont {Giantomassi}, \citenamefont {Bousquet}, \citenamefont
  {Verstraete}, \citenamefont {Hamann}, \citenamefont {Gonze},\ and\
  \citenamefont {Rignanese}}]{vanSetten2018pseudodojo}%
  \BibitemOpen
  \bibfield  {author} {\bibinfo {author} {\bibfnamefont {M.~J.}\ \bibnamefont
  {van Setten}}, \bibinfo {author} {\bibfnamefont {M.}~\bibnamefont
  {Giantomassi}}, \bibinfo {author} {\bibfnamefont {E.}~\bibnamefont
  {Bousquet}}, \bibinfo {author} {\bibfnamefont {M.~J.}\ \bibnamefont
  {Verstraete}}, \bibinfo {author} {\bibfnamefont {D.~R.}\ \bibnamefont
  {Hamann}}, \bibinfo {author} {\bibfnamefont {X.}~\bibnamefont {Gonze}},\ and\
  \bibinfo {author} {\bibfnamefont {G.-M.}\ \bibnamefont {Rignanese}},\
  }\bibfield  {title} {\bibinfo {title} {The {PseudoDojo}: {Training} and
  grading a 85 element optimized norm-conserving pseudopotential table},\
  }\href {https://doi.org/10.1016/j.cpc.2018.01.012} {\bibfield  {journal}
  {\bibinfo  {journal} {Comput. Phys. Commun.}\ }\textbf {\bibinfo {volume}
  {226}},\ \bibinfo {pages} {39} (\bibinfo {year} {2018})}\BibitemShut
  {NoStop}%
\bibitem [{\citenamefont {Mostofi}\ \emph {et~al.}(2008)\citenamefont
  {Mostofi}, \citenamefont {Yates}, \citenamefont {Lee}, \citenamefont {Souza},
  \citenamefont {Vanderbilt},\ and\ \citenamefont
  {Marzari}}]{Mostofi2008Wannier90}%
  \BibitemOpen
  \bibfield  {author} {\bibinfo {author} {\bibfnamefont {A.~A.}\ \bibnamefont
  {Mostofi}}, \bibinfo {author} {\bibfnamefont {J.~R.}\ \bibnamefont {Yates}},
  \bibinfo {author} {\bibfnamefont {Y.-S.}\ \bibnamefont {Lee}}, \bibinfo
  {author} {\bibfnamefont {I.}~\bibnamefont {Souza}}, \bibinfo {author}
  {\bibfnamefont {D.}~\bibnamefont {Vanderbilt}},\ and\ \bibinfo {author}
  {\bibfnamefont {N.}~\bibnamefont {Marzari}},\ }\bibfield  {title} {\bibinfo
  {title} {{Wannier90}: A tool for obtaining maximally-localised wannier
  functions},\ }\href {https://doi.org/10.1016/j.cpc.2007.11.016} {\bibfield
  {journal} {\bibinfo  {journal} {Comput. Phys. Commun.}\ }\textbf {\bibinfo
  {volume} {178}},\ \bibinfo {pages} {685} (\bibinfo {year}
  {2008})}\BibitemShut {NoStop}%
\bibitem [{\citenamefont {Gonze}\ \emph {et~al.}(2020)\citenamefont {Gonze},
  \citenamefont {Amadon}, \citenamefont {Antonius}, \citenamefont {Arnardi},
  \citenamefont {Baguet} \emph {et~al.}}]{Gonze2020}%
  \BibitemOpen
  \bibfield  {author} {\bibinfo {author} {\bibfnamefont {X.}~\bibnamefont
  {Gonze}}, \bibinfo {author} {\bibfnamefont {B.}~\bibnamefont {Amadon}},
  \bibinfo {author} {\bibfnamefont {G.}~\bibnamefont {Antonius}}, \bibinfo
  {author} {\bibfnamefont {F.}~\bibnamefont {Arnardi}}, \bibinfo {author}
  {\bibfnamefont {L.}~\bibnamefont {Baguet}}, \emph {et~al.},\ }\bibfield
  {title} {\bibinfo {title} {The {Abinit project}: {Impact, environment and
  recent developments}},\ }\href
  {https://doi.org/https://doi.org/10.1016/j.cpc.2019.107042} {\bibfield
  {journal} {\bibinfo  {journal} {Comput. Phys. Commun.}\ }\textbf {\bibinfo
  {volume} {248}},\ \bibinfo {pages} {107042} (\bibinfo {year}
  {2020})}\BibitemShut {NoStop}%
\bibitem [{\citenamefont {Aroyo}\ \emph {et~al.}(2006)\citenamefont {Aroyo},
  \citenamefont {Kirov}, \citenamefont {Capillas}, \citenamefont {Perez-Mato},\
  and\ \citenamefont {Wondratschek}}]{Aroyo2006Bilbao}%
  \BibitemOpen
  \bibfield  {author} {\bibinfo {author} {\bibfnamefont {M.~I.}\ \bibnamefont
  {Aroyo}}, \bibinfo {author} {\bibfnamefont {A.}~\bibnamefont {Kirov}},
  \bibinfo {author} {\bibfnamefont {C.}~\bibnamefont {Capillas}}, \bibinfo
  {author} {\bibfnamefont {J.~M.}\ \bibnamefont {Perez-Mato}},\ and\ \bibinfo
  {author} {\bibfnamefont {H.}~\bibnamefont {Wondratschek}},\ }\bibfield
  {title} {\bibinfo {title} {{Bilbao Crystallographic Server}. {II}.
  {Representations} of crystallographic point groups and space groups},\ }\href
  {https://doi.org/10.1107/S0108767305040286} {\bibfield  {journal} {\bibinfo
  {journal} {Acta Crystallogr. Sect. A}\ }\textbf {\bibinfo {volume} {62}},\
  \bibinfo {pages} {115} (\bibinfo {year} {2006})}\BibitemShut {NoStop}%
\bibitem [{\citenamefont {Elcoro}\ \emph {et~al.}(2017)\citenamefont {Elcoro},
  \citenamefont {Bradlyn}, \citenamefont {Wang}, \citenamefont {Vergniory},
  \citenamefont {Cano}, \citenamefont {Felser}, \citenamefont {Bernevig},
  \citenamefont {Orobengoa}, \citenamefont {de~la Flor},\ and\ \citenamefont
  {Aroyo}}]{Elcoro2017Double}%
  \BibitemOpen
  \bibfield  {author} {\bibinfo {author} {\bibfnamefont {L.}~\bibnamefont
  {Elcoro}}, \bibinfo {author} {\bibfnamefont {B.}~\bibnamefont {Bradlyn}},
  \bibinfo {author} {\bibfnamefont {Z.}~\bibnamefont {Wang}}, \bibinfo {author}
  {\bibfnamefont {M.~G.}\ \bibnamefont {Vergniory}}, \bibinfo {author}
  {\bibfnamefont {J.}~\bibnamefont {Cano}}, \bibinfo {author} {\bibfnamefont
  {C.}~\bibnamefont {Felser}}, \bibinfo {author} {\bibfnamefont {B.~A.}\
  \bibnamefont {Bernevig}}, \bibinfo {author} {\bibfnamefont {D.}~\bibnamefont
  {Orobengoa}}, \bibinfo {author} {\bibfnamefont {G.}~\bibnamefont {de~la
  Flor}},\ and\ \bibinfo {author} {\bibfnamefont {M.~I.}\ \bibnamefont
  {Aroyo}},\ }\bibfield  {title} {\bibinfo {title} {Double crystallographic
  groups and their representations on the {Bilbao Crystallographic Server}},\
  }\href {https://doi.org/10.1107/S1600576717011712} {\bibfield  {journal}
  {\bibinfo  {journal} {J. Appl. Crystallogr.}\ }\textbf {\bibinfo {volume}
  {50}},\ \bibinfo {pages} {1457} (\bibinfo {year} {2017})}\BibitemShut
  {NoStop}%
\bibitem [{\citenamefont {King-Smith}\ and\ \citenamefont
  {Vanderbilt}(1993)}]{KingSmith1993Polarization}%
  \BibitemOpen
  \bibfield  {author} {\bibinfo {author} {\bibfnamefont {R.~D.}\ \bibnamefont
  {King-Smith}}\ and\ \bibinfo {author} {\bibfnamefont {D.}~\bibnamefont
  {Vanderbilt}},\ }\bibfield  {title} {\bibinfo {title} {Theory of polarization
  of crystalline solids},\ }\href {https://doi.org/10.1103/PhysRevB.47.1651}
  {\bibfield  {journal} {\bibinfo  {journal} {Phys. Rev. B}\ }\textbf {\bibinfo
  {volume} {47}},\ \bibinfo {pages} {1651} (\bibinfo {year}
  {1993})}\BibitemShut {NoStop}%
\end{thebibliography}%
